\documentclass[prd,twocolumn,preprintnumbers,superscriptaddress,nofootinbib, 10pt]{revtex4-2}
\usepackage{epsfig}
\usepackage{amsmath}
\usepackage{amsfonts}
\usepackage{float}
\usepackage{amssymb}
\usepackage{slashed}
\usepackage{color}
\usepackage{xcolor}
\usepackage{pbox}
\usepackage{subfigure}
\usepackage[hidelinks]{hyperref}
\usepackage{tabularx}
\usepackage{titlesec}
\usepackage{scrextend}
\usepackage[version=3]{mhchem}
\usepackage{orcidlink}

\def\lsim{\lesssim}
\def\gsim{\gtrsim}

\def\be{\begin{eqnarray}}
\def\ee{\end{eqnarray}}

\def\nn{\nonumber}

\def\del{\partial}

\def\D{\text{D}}

\def\bal#1\eal{\begin{align}#1\end{align}}
\def\beq{\begin{equation}}
\def\eeq{\end{equation}}

\def\MeV{{\rm MeV}}
\def\keV{{\rm keV}}

\def\H{\text{H}}
\def\T{\text{T}}
\def\He{{}^4\text{He}}
\def\Het{{}^3\text{He}}
\def\Li{{}^7\text{Li}}
\def\Be{{}^7\text{Be}}
\def\Lisix{{}^6\text{Li}}

\def\tnu{{\nu_{\rm nt}}}
\def\tanu{{\nu_{\rm nt, \alpha}}}
\def\bnu{{\nu_{\rm bg}}}

\def\nud{{\nu {\rm d}}}

\def\Eg{E_\gamma}
\def\Ee{E_{e^-}}
\def\Ep{E_{e^+}}
\def\gnt{{\gamma_{\rm nt}}}
\def\ent{e^-_{\rm nt}}
\def\epnt{e^+_{\rm nt}}

\allowdisplaybreaks[2]

\begin{document}

\title{Bipartite Solution to the Lithium Problem}

\author{Sougata Ganguly\,\orcidlink{0000-0002-8742-0870}}
\email{sganguly0205@ibs.re.kr}
\affiliation{Particle Theory  and Cosmology Group (PTC),
Center for Theoretical Physics of the Universe (CTPU), \\
Institute for Basic Science, Daejeon 34126, Republic of Korea}

\author{Tae Hyun Jung\,\orcidlink{0000-0002-2515-9283}}
\email{thjung0720@gmail.com}
\affiliation{Particle Theory  and Cosmology Group (PTC),
Center for Theoretical Physics of the Universe (CTPU), \\
Institute for Basic Science, Daejeon 34126, Republic of Korea}

\author{Tae-Sun Park\,\orcidlink{0000-0001-8984-2922}}
\email{tspark@ibs.re.kr}
\affiliation{Center for Exotic Nuclear Studies (CENS), \\
Institute for Basic Science, Daejeon 34126, Republic of Korea}

\author{Chang Sub Shin\,\orcidlink{0000-0002-4211-5653}}
\email{csshin@cnu.ac.kr}
\affiliation{Department of Physics and Institute for Sciences of the Universe, \\
Chungnam National University, Daejeon 34134, Republic of Korea}
\affiliation{Particle Theory  and Cosmology Group (PTC), 
Center for Theoretical Physics of the Universe (CTPU), \\
Institute for Basic Science, Daejeon 34126, Republic of Korea}
\affiliation{Korea Institute for Advanced Study, Seoul 02455, South Korea}

\preprint{CTPU-PTC-26-11}

\begin{abstract}
The primordial lithium problem remains a persistent motivation for new-physics modifications of Big Bang nucleosynthesis, yet the precision of the observed deuterium abundance now places strong constraints on such attempts. This indicates that the challenge is not simply to reduce $^{7}\mathrm{Li}$, but to realize the correlated shifts among light-element abundances required to do so without spoiling deuterium. We investigate this issue in a concrete two-step decay scenario involving two unstable particles undergoing sequential late decays. In the first stage, a majoron with lifetime $\tau_J \sim 10\,\text{--}\,10^4\,\mathrm{sec}$ decays predominantly into neutrinos, increasing the neutron abundance and thereby reducing the primordial $^{7}\mathrm{Li}+\!{}^{7}\mathrm{Be}$ yield. This mechanism, however, simultaneously drives deuterium above the observationally allowed range. In the second stage, an axion-like particle with a longer lifetime $\tau_\phi \gtrsim 10^5\,\mathrm{sec}$ decays into photons, inducing late-time photodissociation that compensates the excess deuterium without erasing the earlier reduction of lithium, while further amplifying the depletion of $^{7}\mathrm{Li}+\!{}^{7}\mathrm{Be}$. Although the setup is model-dependent, it serves as an explicit proof of concept that the lithium abundance can be lowered consistently with current deuterium constraints. More broadly, our analysis highlights that a viable resolution may require a nontrivial combination of decay channels and decay epochs, and clarifies the pattern of abundance response that successful late-decay scenarios must achieve.
\end{abstract}

\maketitle

\section{INTRODUCTION}

The discrepancy in $\Li$ abundance between the predicted value of Big Bang nucleosynthesis (BBN)\,\cite{Wagoner:1966pv} and the observed value based on the so-called Spite plateau\,\cite{Spite:1982dd} has been a long-standing puzzle, known as the lithium problem.
The standard BBN (SBBN) predicts $(\Li/\H)_{\rm SBBN} = (5.464\pm0.220)\times 10^{-10}$\,\cite{Pitrou:2018cgg, Pitrou:2020etk} (see also Refs.\,\cite{Yeh:2022heq, Pisanti:2007hk, Gariazzo:2021iiu, Burns:2023sgx} where theoretically predicted numbers are consistent with a small difference)
which is almost three times higher than its observed value $(\Li/\H)_{\rm obs} = (1.45 \pm 0.25)\times 10^{-10}$\,\cite{pinto2021metal,ParticleDataGroup:2024cfk} 
(roughly $4\sigma$ deviation). 
Since the main nuclear reactions in SBBN are well measured and the baryon density is fixed by the CMB, this tension suggests that there may be either some new astrophysical effects or physics beyond the Standard Model (BSM).

One possible solution is that the primordial $\Li$ has been depleted during stellar evolution, so the observed abundance today may be lower than the original value.
If the depletion is large enough, the lithium problem can be solved.
This idea, often called the stellar depletion scenario, has been studied in many works 
(see, for example, Refs.\,\cite{1984ApJ...282..206M, 1988ApJ...335..971V, 1991ApJ...371..584P, 1994ApJ...433..510C, 1995A&A...295..715V, 1998ApJ...502..372V, 2001A&A...376..955S, Richard:2001qp, 2002ApJ...580.1100R, Richard:2004pj, Piau:2006sw, Korn:2006tv, 2008ApJ...689.1279P, 2013A&A...552A.131V, 2015MNRAS.452.3256F, 2019MNRAS.489.3539G, 2020A&A...633A..23D, 2020A&A...638A..81T, 2021A&A...654A..46D, 2021A&A...646A..48D, 2024A&A...690A.245B}).
Although the idea is simple and plausible, reproducing the flat Spite plateau and its small dispersion over a wide range of metallicity and temperature has been challenging.
Various macroscopic flows involving different physics turned out to be important, and some models incorporating them could successfully reproduce the Spite plateau\,\cite{Korn:2024gel}.
Moreover, the recent non-observation of $^6{\rm Li}$ in three Spite plateau stars\,\cite{2022MNRAS.509.1521W} 
indicates a large depletion of $^6{\rm Li}$, supporting the stellar depletion of $\Li$\,\cite{Fields:2022mpw}.
Nevertheless, there are still model parameters that lack derivation from the first principles, and thus we still need an explanation of why these model parameters should be specific values that solve the $\Li$ problem.

Alternatively, the primordial $\Li$ abundance may have been reduced by BSM effects, which we focus on in this work.
In SBBN, most of the final $\Li$ comes from $\Be$ that later decays into $\Li$ after BBN.
Thus, the lithium problem indicates that the $\Be$ abundance from SBBN is too large, and various BSM scenarios have been proposed to reduce it\,\cite{
Reno:1987qw, Jedamzik:2004er, Pospelov:2010cw, AlbornozVasquez:2012emy, Coc:2014gia, Kusakabe:2014ola, Kawasaki:2017bqm, 
Kusakabe:2013sna, Salvati:2016jng, Chang:2024mvg,
Goudelis:2015wpa, 
Erken:2011vv,Kusakabe:2012ds, 
Kawasaki:2020qxm,
Luo:2018nth, Yamanaka:2025xyv,
Koren:2022axd}. 

However, the difficulty lies in maintaining the prediction for the primordial abundances of $\D$ and $\He$ whose currently observed values
are $(25.08 \pm 0.29) \times 10^{-6}$, and $0.245 \pm 0.003$, respectively
\cite{ParticleDataGroup:2024cfk}.
For example, if a BSM scenario produces additional neutrons, the $\Be$ abundance can be reduced via $\Be (n, p) \Li$, and the produced $\Li$ gets subsequently destroyed by $\Li (p,\,\alpha) \He$. 
As primordial $\Be$ decays to $\Li$ via the electron capture after recombination with the lifetime $\sim 5\times 10^6\,\sec$, the current $\Li$ (which is primordial $\Li+\!\Be$) can be reduced through this mechanism
\,\cite{Reno:1987qw, Jedamzik:2004er, Pospelov:2010cw, AlbornozVasquez:2012emy, Coc:2014gia, Kusakabe:2014ola, Kawasaki:2017bqm, 
Kusakabe:2013sna, Salvati:2016jng, Chang:2024mvg}.
There are various ways to induce the excess neutrons: the direct decay of heavy particles, or indirectly through $p\to n$ conversion induced by injection of other particles such as mesons or neutrinos.
However, this scenario inevitably increases the $p(n, \gamma) \D$ rate due to the more neutrons, leading to an overproduction of $\D/\H$, and is thus ruled out after the precise measurement of $\D/\H$ in \cite{Cooke:2013cba, Cooke:2016rky, 
Riemer-Sorensen:2014aoa, Balashev:2015hoe, Riemer-Sorensen:2017pey, Zavarygin:2017cov, Cooke:2017cwo} .
Similarly, photon injection at a late time can lead to photo-dissociation of $\Be$ and $\Li$.
However, the injected photons also destroy deuterium too much unless the energy of the injected photon lies in a narrow range between $1.59\, \rm MeV$ and $2.22 \,\rm MeV$ (between $\D$ and $\Be$ photo-dissociation thresholds)\,\cite{Kawasaki:2020qxm}.
Thus, most of the parameter space in these simple solutions is already in severe tension with deuterium\footnote{
Including the theoretical uncertainty of $\D/\H$, which is roughly $5\,\%$\,\cite{Yeh:2020mgl}, can, in principle, mitigate the difficulty, but does not rescue these solutions.}.

\begin{figure}
    \centering
    \includegraphics[width=0.48\textwidth]{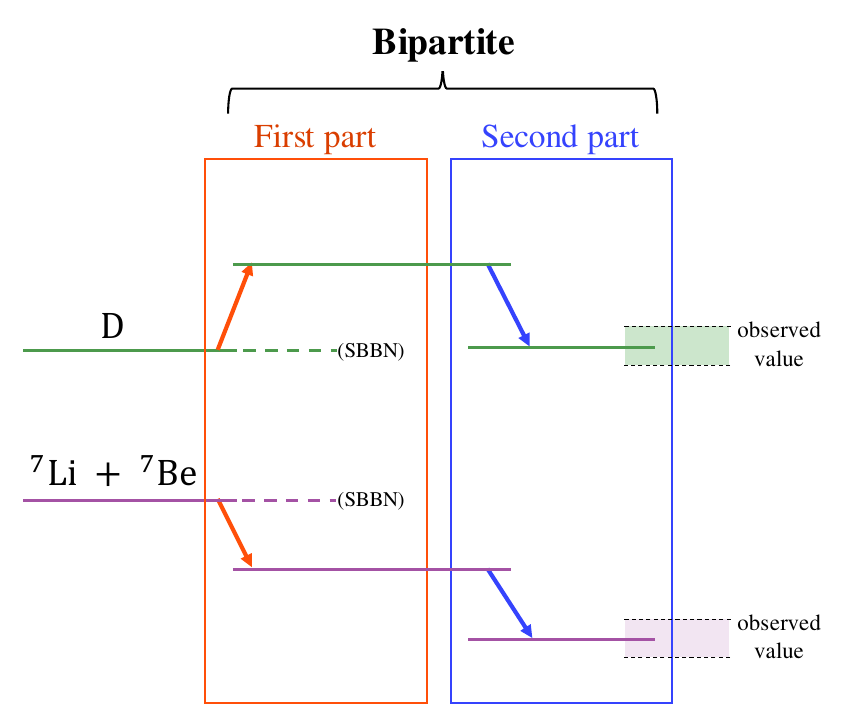}
    \caption{Schematic picture of the bipartite solution.}
    \label{fig:schematic}
\end{figure}

This apparent difficulty may stem from our tendency to favor simple or natural explanations, leading us to overlook possibilities that rely on accidental coincidences.
In this work, we consider a solution composed of two distinct scenarios, which we call \emph{bipartite solution} (see Fig.\,\ref{fig:schematic} for a schematic picture).
The first part of our bipartite solution increases $\D$ and decreases $\Li+\!\Be$, while the second part decreases both $\D$ and $\Li+\!\Be$.
Consequently, the effects on the $\D$ abundance from the two parts cancel each other, while $\Li+\!\Be$ can be reduced.
As one can easily expect, this scenario requires tuning the abundances of two different particles so that the $\D$ abundance remains unchanged, and the goal of this paper is to quantify the degree of precision required for this balance.
We do not include the theoretical uncertainty of $\D/\H$ in our analysis, leading to a conservative estimation of the degree of tuning.

As a concrete example, we consider two late-time decaying particles: a majoron decaying into neutrinos, and an axion-like particle (ALP) decaying into photons.
The majoron $J$ couples with the standard model neutrinos via 
$-\dfrac{g}{2} J \bar{\nu}_\alpha \gamma_5 \nu_\alpha$ where $\alpha = e, \mu, \tau$, 
assuming the flavor universality as an approximation. 
The decay of $J$ injects energetic neutrinos, and these neutrinos enhance the abundance of neutrons via $p \to n$ conversion. 
As discussed earlier, these excess neutrons can destroy $\Be$, and explain the observed $\Li$ abundance, but at the same time, $\D$ will be overproduced~\cite{Chang:2024mvg}.
As the second decaying particle, an ALP $\phi$ couples with a pair of photons via 
$-\dfrac{g_{a \gamma \gamma}}{4}\phi F^{\mu \nu} \tilde{F}_{\mu \nu}$. 
The photons produced from the $\phi$ decay destroy both $\D$ and $\Li+\!\Be$, bringing $\D$ back to the observed value while enhancing the depletion of the final $\Li$ abundance.

In this work, we take the initial yields, $Y_J^{(0)}$ for $J$ and $Y_\phi^{(0)}$ for $\phi$, as free parameters, since they strongly depend on the reheating temperature and their UV completion.
We also restrict the majoron mass $m_J \sim 100\,\MeV$ for simplicity (neutrinos from a heavier majoron can produce muons, pions, etc.), and the ALP mass $m_\phi \sim 20\,\MeV$, which leads to the injected photon energy lying above the $\D$ threshold $\sim 2\,\MeV$ but below the $\He$ threshold $\sim 20\,\MeV$.
If the photon energy were above the $\He$ threshold, the photo-dissociation of $\He$ leads to enhancing the $\D$ abundance, disabling the cancellation required for the observed $\D/\H$. In this case, our bipartite solution would not work.

For the majoron part to work properly, we need its lifetime $\tau_J$ to be around/after the deuterium bottleneck, $10\,\sec \lsim \tau_J \lsim 10^4\,\sec$\,\cite{Chang:2024mvg}.
On the other hand, the lifetime of $\phi$ has to be greater than $10^5\,\sec$ to make injected photons efficiently destroy $\D$ and $\Li$ before they get thermalized\,\cite{Kawasaki:1994sc,Hufnagel:2018bjp}.
Therefore, we have a hierarchy $\tau_J\ll \tau_\phi$, which indeed cleanly separates our analysis into two parts: (i) BBN evolution with $J$, and (ii) photo-dissociation driven by $\phi$.

In part (i), we follow Ref.\,\cite{Chang:2024mvg}, ignoring the contribution from $\phi$, which is a good approximation as long as $\phi$'s initial abundance is sufficiently small (we check the consistency later on).
This is reviewed in Sec.\,\ref{sec:Majoron_decay}.
We then take the boundary conditions obtained in part (i) and estimate the photo-dissociation processes in part (ii) mainly following Ref.\,\cite{Cyburt:2002uv}, of which details are reviewed in Sec.\,\ref{sec:ALP_decay}.
We combine these two parts and present our results in Sec.\,\ref{sec:results}, and conclude in Sec.\,\ref{sec:conclu}.

\section{EFFECTS OF PARTICLE INJECTION ON BBN}
\label{sec:effect_on_bbn}

Decay products of a long-lived particle in BSM interact with light nuclei and also modify the background plasma evolution.
This alters the prediction of SBBN, and the detailed effect depends on what particles are injected from the decay. 
Energetic neutrino injection alters the neutron-proton freeze-out process, or directly interacts with nuclei depending on the lifetime of the BSM particle\,\cite{1984MNRAS.210..359S, Chang:1993yp, Kanzaki:2007pd, Kusakabe:2013sna, Salvati:2016jng, Chang:2024mvg, Bianco:2025boy}.
If the decay products are $e^\pm$ or $\gamma$, energetic photons at around or after $10^5\,\rm sec$ can photo-dissociate $\D$, $\He$, and $\Het$\,\cite{1988ApJ...331...33S, Sarkar:1984tt, Ellis:1984er, Kawasaki:1994sc, Protheroe:1994dt, Holtmann:1998gd, Kawasaki:2000qr,Cyburt:2002uv, Jedamzik:2006xz, Poulin:2015woa, Poulin:2015opa, Hufnagel:2018bjp, Forestell:2018txr}.
As discussed in Refs.\,\cite{Reno:1987qw, Kohri:1999ex, Kawasaki:2000en, Kohri:2001jx, Kawasaki:2004qu, Jedamzik:2004er, Kawasaki:2004fw, Kawasaki:2004yh, Kawasaki:2004qu, Kohri:2005wn, Jedamzik:2006xz, Kawasaki:2008qe, Cyburt:2009pg, Jedamzik:2009uy, Pospelov:2010cw, Cyburt:2010vz, Henning:2012rm, Fradette:2014sza, Berger:2016vxi, Kawasaki:2017bqm, Fradette:2017sdd,Hasegawa:2019jsa, Boyarsky:2020dzc, Chen:2024cla, Omar:2025jue, Angel:2025dkw, Jung:2025dyo}, 
hadronic injections can significantly modify the predictions of SBBN.
For the purpose of our study, we only discuss two specific scenarios: i) neutrino injection from majoron decay, and ii) photon injection from the ALP decay. 

Modified evolution of a nucleus $A=p,\,n,\,\D,\,\cdots$ can be obtained by solving
\bal
 \frac{d X_A}{dt} = 
 \Gamma_A^{\rm SBBN}
 - \sum_{B\neq A} \Big[ \delta \Gamma_{A\to B} X_A - \delta \Gamma_{B\to A} X_B\Big],
 \label{eq:BE_X_J}
\eal
where $X_A = n_A/n_b$ is the ratio of $A$ number density $n_A$ and the total baryon number density $n_b$, and $\Gamma_A^{\rm SBBN}$ denotes the production rate of $A$ in 
SBBN, which is a function of the abundances of other species.

The second term on the right-hand side (RHS) of Eq.\,\eqref{eq:BE_X_J} corresponds to the BSM effects.
For $\delta \Gamma_{A\to B}$ and $\delta \Gamma_{B\to A}$, it is crucial to obtain non-thermal distributions of injected neutrinos $\tnu$ or photons $\gnt$, which we denote $f_\tnu$ and $f_\gnt$, respectively, as well as modified background plasma properties such as background neutrino temperature.
In the following subsections, we describe the derivation of these functions and examine their impact on the background plasma evolution.

\label{sec:LLP_BBN}

\subsection{Part I: early-time neutrino injection}
\label{sec:Majoron_decay}
The decay of majoron produces energetic neutrinos with initial energy $m_J/2$.
The neutrino momentum distribution function $f_{\tanu}$ for a flavor $\alpha=e,\,\mu,\,\tau$ can be obtained by solving the Boltzmann equation
\bal
\dfrac{\del f_{\tanu}}{\del t} - H p \dfrac{\del f_{\tanu}}{\del p}
 = \dfrac{2 \pi^2  n_J}{3 E^2 \tau_J} \delta \left(E - \dfrac{m_J}{2}\right)
- {\cal C}_{\tanu}\,,
\label{eq:BE_fnu}
\eal
where $H$ is the Hubble expansion rate and 
$p \equiv |\vec{p}|$ and $E$ are the magnitude of the three-momentum and energy of $\tanu$, respectively.
The first term on the  RHS
corresponds to the source term of $\tanu$ from the majoron decay, with its number density denoted by $n_J$. 
For simplicity, we approximate $n_J (T) = Y_J^{(0)} s(T) \exp(-t/\tau_J)$ where $s(T)$ is the entropy density of the background plasma.

The last term of Eq.\,\eqref{eq:BE_fnu} generates the reduction of the number of $f_\tanu$ due to its scattering with the background electrons and neutrinos, ignoring their asymmetry and the contribution from the baryons. 
The explicit form of ${\cal C}_{\tanu a \to bc}$ is written as
\bal
{\cal C}_{\tanu} = 
\sum_{a,b,c}
\dfrac{S f_{\nu_{\tanu}}}{2E} 
\int d \Pi_a & d \Pi_b d \Pi_c |{\cal M}_{\tanu a \to bc}|^2 f_a
\nn \\
&(2 \pi)^4 \delta^4 (P + P_a - P_b - P_c) \,.
\label{eq:damping_nu}
\eal
Here $a$, $b$, and $c$ denote the background particles $e^+$, $e^-$ $\nu$ or $\bar \nu$ with their four momenta $P_a$, $P_b$, and $P_c$. 
${\cal M}_{\tanu a \to bc}$ is matrix amplitude of the process $\tanu a \to bc$ and $S$ is its symmetry factor. 
$d \Pi_i = d^3\vec{p}_i/(2 \pi)^3 2 E_i$ is the Lorentz-invariant phase space measure and $f_a$ is the momentum distribution of the initial state particle $a$. 
The explicit form of the matrix amplitude for different processes can be found in Appendix A of Ref.\,\cite{Chang:2024mvg}.

In Eq.\,\eqref{eq:BE_fnu}, we dropped positive scattering terms that come from the elastic scattering processes as an approximation.
This is equivalent to assuming that a non-thermal neutrino gets thermalized by a single scattering, which leads to an underestimated number of $\tanu$ in the intermediate energy range.
However, for $\tanu$ to efficiently produce neutrons, which are the key ingredient to reduce $\Li+\!\Be$, we need the collision terms to be small enough, leading us to $\tau_J \gsim 10\,\sec$.
Thus, the positive scattering terms are also suppressed in this range, and thus neglecting them is a good approximation.

$f_\tanu$ modifies the BBN evolution via $\delta \Gamma_{A\to B}$ in Eq.\,\eqref{eq:BE_X_J} with
\bal
\delta \Gamma_{A\to B} &=\dfrac{1}{2 \pi^2} \int dE_{\tnu}\, E_{\tnu}^2 \,f_{\tnu}(\sigma v)_{\tnu A \to B } + \Delta \Gamma^{\rm (weak)}_{A\to B}\,\,,
\label{eq:non_thermal_nu_rate}
\eal
where $\sigma_{\tnu A \to B}$ is the scattering cross section of $\tnu$ and $A$ which produces $B$.
The last term $\Gamma^{\rm (weak)}_{A\to B}$ accounts for the effect coming from the modified temperature of the background neutrinos, which can affect $n\leftrightarrow p$ conversion processes. 
We include this effect although it is negligible for $\tau_J \gsim 10\,\sec$\,\cite{Chang:2024mvg}.

In addition, non-thermal neutrinos transfer part of their energy to the $e\gamma B$ sector through scatterings with the background plasma, and this effect is taken into account in the evolution of $\rho_{e\gamma B}$ through the modified Boltzmann equation 
\bal
&\dot \rho_{e\gamma B} + 3H(\rho_{e\gamma B} + P_{e\gamma B}) 
= -T^4 H(T) ({\cal N}(T) +\Delta {\cal N}(T))\,\,,
\label{eq:rho_e_gam_B}
\eal
where $P_{e \gamma B}$ is the pressure of the $e \gamma B$ sector.
The effect of the non-instantaneous neutrino decoupling in SBBN is encoded in ${\cal N} (T)$ term\,\cite{Mangano:2001iu, Mangano:2005cc, Mangano:2006ar} on the RHS and $\Delta {\cal N}(T)$ is its correction in the presence of $\tnu$, which is given by
\bal
\Delta {\cal N}(T)=
\sum_{\alpha} \frac{\Gamma (\tanu \to e)}{H(T)}
\frac{\rho_{\tanu}(T)}{T^4}\, .
\eal
Here $\rho_{\tanu}(T)$ is the energy density of non-thermal neutrinos, and $\Gamma (\tanu \to e)$ is the effective rate for energy transfer from $\tanu$ to the electromagnetic plasma, whose explicit form is given in Appendix A of Ref.\,\cite{Chang:2024mvg}.

Energy transfer from $\tnu$ to $e \gamma B$ sector also dilutes the baryon-to-photon ratio $\eta_B$.
To fix its final value after the BBN epoch as $\eta_{b, \rm fin} = 6.1 \times 10^{-10}$ to agree with CMB fitting\,\cite{Yeh:2022heq}, we modify the initial baryon asymmetry as
\bal
\frac{\eta_{B,{\rm ini}}}{\eta_{B,{\rm fin}}} \! = & \Bigg[
2.73 
\!-\! \dfrac{45}{2\pi^2 {g_*s}(T_f)} 
\!\! \int_{T_{\nud}}^{T_f} \!\!
\dfrac{\sum_{\alpha}\!\Gamma (\tanu \! \to \! e) \rho_{\tanu}\!(T)}{H(T) T^2 T_{\bnu}^3}dT
\Bigg],
\label{Eq:etaB_dilute}
\eal
where $T_\bnu$  is the background neutrino temperature
and we take $T_{\nud} = 2.3\,\rm MeV$ and $T_f = 5\,\rm keV$.

\begin{figure}
    \centering
\includegraphics[width=0.48\textwidth]{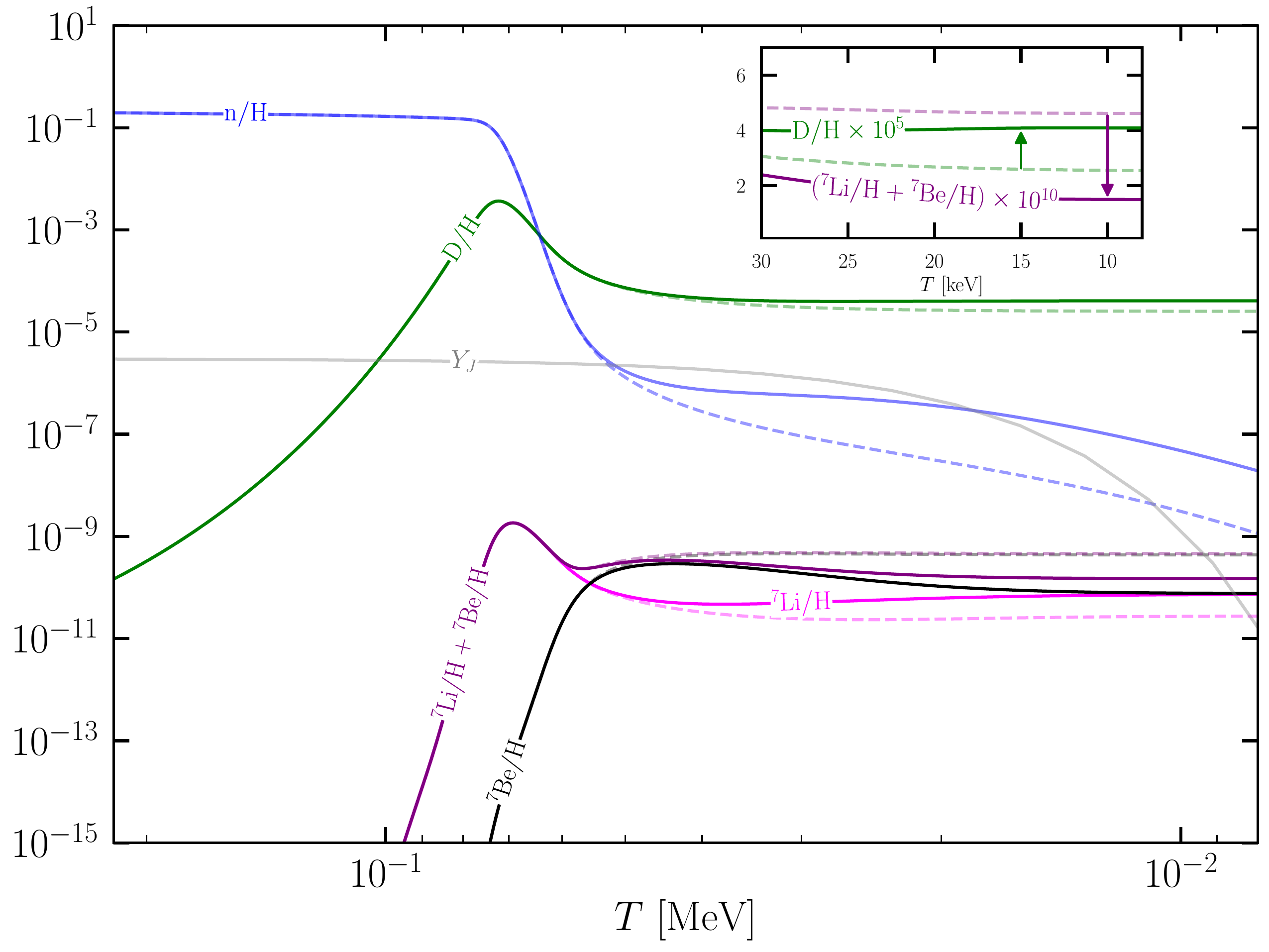}
    \caption{Evolution of 
    $n/\H$ (blue), $\D/\H$ (green), $\Li/\H$ (magenta), $\Be/\H$ (black),  $(\Li/\H + \Be/\H)$ (purple), and 
    $Y_J$ (gray) as a function of $T$ for
    $m_J = 100\,\rm MeV$, 
    $\tau_J = 10^3\,\rm sec$,
    and $Y_J^{(0)} = 3 \times 10^{-6}$. The dashed lines
    depict their evolution in SBBN.
    } 
    \label{fig:evol_wJ}
\end{figure}

In Fig.\,\ref{fig:evol_wJ}, we show the evolution of light element abundances in $T$ for $m_J = 100\,\rm MeV$, $Y_J^{(0)} = 3\times 10^{-6}$, and $\tau_J = 10^3\,\rm sec$. 
Different colored lines correspond to different elements, as indicated by labels in each line, and the dashed lines represent the evolution in SBBN.
As discussed earlier, the $p \to n$ conversion rate is enhanced due to the injection of energetic neutrinos from majoron decay, leading to an enhanced neutron density (see the blue lines).
These additional neutrons subsequently make the deuterium production more efficient through the 
$p(n,\gamma)\D$
reaction, resulting in a larger $\D/\H$ in comparison to its SBBN value (see the green lines). 
On the other hand, the effect of these additional neutrons on $\Li + \Be$ abundance is opposite. 
The excess neutrons dissociate $\Be$ through $\Be(n, p)\Li$, and the produced $\Li$ is subsequently converted into $\He$ via $\Li(p, \He) \He$. 
As a result of the $\Be$ dissociation, the final $\Li + \Be$ abundance (solid purple line) is smaller than its SBBN value (dashed purple line).

\begin{figure}[t]
    \centering
    \includegraphics[width=0.5\textwidth]{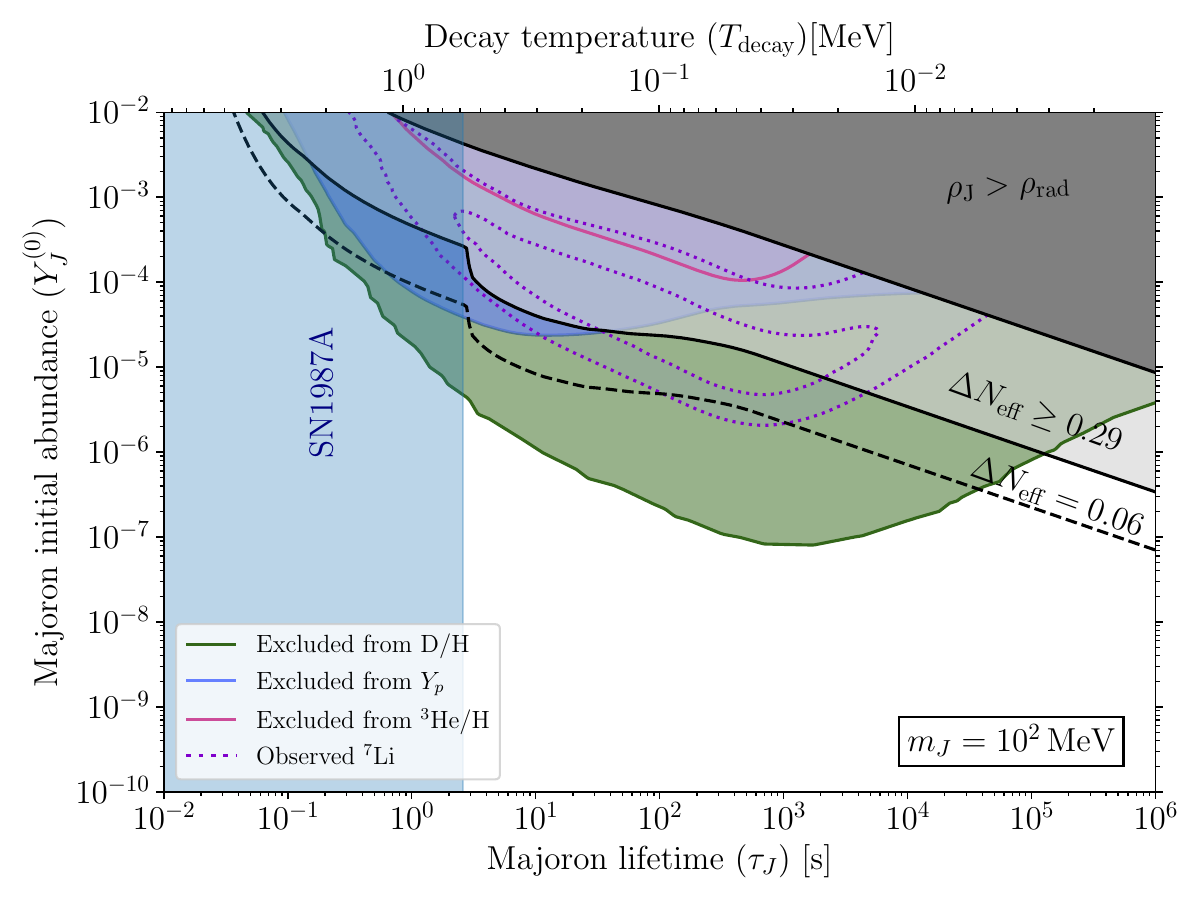}
    \caption{BBN constraint on $\tau_J - Y_J^{(0)} $ plane for $m_J = 100\,\rm MeV$, taken from \cite{Chang:2024mvg}. Exclusion limit
    at 95\% confidence level (C.L.) from $\D$,
    $\He$, and $\Het$ are depicted by green, blue,
    and magenta colored regions, respectively. The dark
    gray region is excluded from the majoron domination
    during/after BBN. The observed value of lithium abundance
    can be explained in the parameter region outlined
    by purple dotted lines. $\Delta N_{\rm eff}$ constraint 
    from Planck 2018 and future experiment CMB-S4\,\cite{CMB-S4:2016ple}
    is depicted by black solid and dashed lines, respectively. The light blue region is excluded from SN 1987A constraint.
    }
    \label{fig:BBN_exclu_nu_injection}
\end{figure}

We depict in Fig.\,\ref{fig:BBN_exclu_nu_injection} (taken from \cite{Chang:2024mvg}) the constraints in $\tau_J -Y_J^{(0)}$ plane for $m_J = 100\,\rm MeV$, assuming that there is no second part.
The constraints from the overabundance of $\D$, $\He$, and $\Het$ correspond to the green, blue, and magenta regions, respectively, and a rough estimation of $\Delta N_{\rm eff}$ constraint is shown by the gray shaded region.
The supernova 1987A constraint from \cite{Fiorillo:2022cdq} is also depicted by the light blue shaded region, where we use the relation $\tau_J \simeq 16\pi/3g^2 m_J$ with an approximation of flavor universal decay. This constraint can be improved if a future galactic supernovae
explosion is observed at Hyper-Kamiokande, as discussed in Ref.\,\cite{Telalovic:2024cot}. 

Importantly, the region surrounded by the purple dotted curves in Fig.\,\ref{fig:BBN_exclu_nu_injection} corresponds to the parameter space where the observed $\Li$ can be explained.
However, as we pointed out in the introduction, this region is already excluded by the overproduction of $\D$.
This will be resolved by combining with the second part, where injected photons reduce both $\D$ and $\Li + \Be$ abundances.

\subsection{Part II: late-time photon injection}
\label{sec:ALP_decay}
In the second part, photons injected from the decay of ALP, each with initial energy $E_0 = m_\phi/2$, scatter off the background photons and charged particles. 
These interactions initiate electromagnetic (EM) cascades, leading to the production of secondary photons and electrons/positrons with lower energies.

Following Ref.\,\cite{Hufnagel:2018bjp}, the momentum distributions of $\gnt$, $\ent$, and $\epnt$ can be obtained by solving the Boltzmann equations,
\bal
\dfrac{\partial \hat f_{\gnt}}{\partial t} - H p \dfrac{\partial \hat f_{\gnt}}{\partial p}
&=  \xi_\gamma \delta \left(E-E_0\right) -\Gamma_\gamma \hat f_{\gnt}
\nn\\
&+\int_E^{E_0} \!\! d \Eg^\prime \hat f_{\gnt}(\Eg^\prime) 
K_{\gamma\to \gamma} (\Eg^\prime, E) 
\nn\\
&+\int_E^{E_0} \!\! d \Ee^\prime \hat f_{\ent}(\Ee^\prime) K_{e^-\to \gamma} (\Ee^\prime, E)\nn\\
&+\int_E^{E_0} \!\! d \Ep^\prime \hat f_{\epnt}(\Ep^\prime) K_{e^+\to \gamma} (\Ep^\prime, E) 
\, ,
\label{eq:f_gamma_full}\\
\dfrac{\partial \hat f_{\ent}}{\partial t} - H p \dfrac{\partial \hat f_{\ent}}{\partial p}
&= -\Gamma_{e^-} \hat f_{\ent} 
\nn \\
&+ \int_E^{E_0} \!\! d \Eg^\prime \hat f_{\gnt}(\Eg^\prime) 
K_{\gamma\to e^-} (\Eg^\prime, E)
\nn\\
&+\int_E^{E_0} \!\! d \Ee^\prime \hat f_{\ent}(\Ee^\prime) K_{e^-\to e^-} (\Ee^\prime, E)
\,,\label{eq:f_e_full}\\
\dfrac{\partial \hat f_{\epnt}}{\partial t} - H p \dfrac{\partial \hat f_{\epnt}}{\partial p}
&= -\Gamma_{e^+} \hat f_{\epnt} 
\nn \\
&+ \int_E^{E_0} \!\! d \Eg^\prime \hat f_{\gnt}(\Eg^\prime) 
K_{\gamma\to e^+} (\Eg^\prime, E)
\nn\\
&+\int_E^{E_0} \!\! d \Ep^\prime \hat f_{\epnt}(\Ep^\prime) K_{e^+\to e^+} (\Ep^\prime, E)
\,,\label{eq:f_e+_full}
\eal
where we denote $\hat f_a \equiv \frac{dn_a}{dE} = g_a \frac{p^2}{2\pi^2} \frac{dp}{dE} f_a$ as energy distribution of species $a=\gnt$, $\ent$ and $\epnt$ with its number of degrees of freedom $g_a$.
$K_{a\to b}(E',E)$ corresponds to differential scattering rate of initial state $a$ with energy $E'$ becoming $b$ with energy $E$, while $\Gamma_a$ is the total reaction rate of $a$.
We take $K_{a\to b}$ and $\Gamma_a$ from Appendix B of Ref.\,\cite{Hufnagel:2018bjp}, where injected photon energy is assumed to be much greater than the background photon energy.
The first term on the RHS of Eq.\,\eqref{eq:f_gamma_full} is the source term with $\xi_\gamma = 2 n_\phi(t)/\tau_\phi$.
We approximate the $\phi$ number density $n_\phi(t) = Y_\phi^{(0)} s(T) \exp(-t/\tau_\phi)$, taking its initial yield $Y_\phi^{(0)}$ and lifetime $\tau_\phi$ free parameters.

Since the time scale of the scattering processes is much shorter than the Hubble time scale\,\cite{Kawasaki:1994sc}, we can take stationary solutions which can be obtained by 
taking $\partial f_a/\partial t = 0$ in Eq.\,\eqref{eq:f_gamma_full}\,--\,\eqref{eq:f_e+_full};
\bal
\Gamma_{\gnt} (E)F_{\gnt}(E) &= \dfrac{\xi_\gamma K_{\gamma \to \gamma} (E^0,E)}{\Gamma_\gamma(E_0)} 
\nn\\
&+\int_E^{E_0} d \Eg^\prime K_{\gamma \to \gamma} (\Eg^\prime, E) F_{\gnt} (\Eg^\prime)
\nn\\
&+ \int_E^{E_0} d \Ee^\prime K_{e^- \to \gamma} (\Ee^\prime, E) F_{\ent} (\Ee^\prime) 
\nn\\
&+ \int_E^{E_0} d \Ep^\prime K_{e^+ \to \gamma} (\Ep^\prime, E) F_{\epnt} (\Ep^\prime) 
\label{eq:f_gamma_reduced}\,,\\
\Gamma_{e^-} (E)F_{\ent}(E) &= \dfrac{\xi_\gamma K_{\gamma \to e^-} (E_0,E)}{\Gamma_\gamma(E_0)}
\nn\\
&+ \int_E^{E_0} d \Eg^\prime K_{\gamma \to e^-} (\Eg^\prime, E) F_{\gnt} (\Eg^\prime)
\nn\\
&+\int_E^{E_0} d \Ee^\prime K_{e^- \to e^-} (\Ee^\prime, E) F_{\ent} (\Ee^\prime)\label{eq:f_el_reduced}\,,\\
\Gamma_{e^+} (E)F_{\epnt}(E) &= \dfrac{\xi_\gamma K_{\gamma \to e^+} (E_0,E)}{\Gamma_\gamma(E_0)}
\nn\\
&+ \int_E^{E_0} d \Eg^\prime K_{\gamma \to e^+} (\Eg^\prime, E) F_{\gnt} (\Eg^\prime)
\nn\\
&+ \int_E^{E_0} d \Ep^\prime K_{e^+ \to e^+} 
(\Ep^\prime, E)  F_{\epnt} (\Ep^\prime) 
\label{eq:f_ps_reduced} , 
\eal
where we define
\bal
&F_{\gnt} (E) = \hat f_{\gnt} - \dfrac{\xi_\gamma \delta (E-E_0)}{\Gamma_{\gnt}(E)} \,,\nn\\
&F_{\ent} (E) = \hat f_{\ent} \,,\nn\\
&F_{\epnt} (E) = \hat f_{\epnt}\,.
\eal
To obtain $F_a$ numerically, we discretize the energy interval $E_{\rm min}\le \Eg$, $\Ee$, and $\Ep\le E_0$ into $N$ bins, and write the 
$F_{\gnt}(E_i)$, $F_{\ent} (E_i)$, and $F_{\epnt} (E_i)$ for $0 \leq i \leq N$ as
\bal
F_{\gnt} (E_i)& = \dfrac{1}{\Gamma_{\gnt}(E_i)} \Bigg[
\dfrac{\xi_\gamma K_{\gamma \to \gamma} (E_0,E_i)}{\Gamma_{\gnt}(E^0)} 
\nn \\
&\quad +\sum_{E_j>E_i}^{E_N = E_0} \Delta E_j K_{\gamma  \to \gamma} (E_j, E_i) F_{\gnt} (E_j)
\nn\\
&\quad + \sum_{E_j>E_i}^{E_N = E_0} \Delta E_j K_{e^-  \to \gamma} (E_j, E_i) F_{\ent} (E_j)
\nn\\
&\quad + \sum_{E_j>E_i}^{E_N = E_0} \Delta E_j K_{e^+  \to \gamma} (E_j, E_i) F_{\epnt} (E_j)
\Bigg], \\
F_{\ent} (E_i)& = \dfrac{1}{\Gamma_{\ent}(E_i)} \Bigg[
\dfrac{\xi_\gamma K_{\gamma \to e^-} (E_0,E_i)}{\Gamma_{\gnt}(E^0)}
\nn\\
&\quad +\sum_{E_j>E_i}^{E_N = E_0} \Delta E_j K_{\gamma  \to e^-} (E_j, E_i) F_{\gnt} (E_j)
\nn\\
&\quad +\sum_{E_j>E_i}^{E_N = E_0} \Delta E_j K_{e^-  \to e^-} (E_j, E_i) F_{\ent} (E_j)
\Bigg]\,,\nn\\
F_{\epnt} (E_i)& = \dfrac{1}{\Gamma_{\epnt}(E_i)} \Bigg[
\dfrac{\xi_\gamma K_{\gamma \to e^+} (E_0,E_i)}{\Gamma_{\gnt}(E^0)}
\nn\\
&\quad +\sum_{E_j>E_i}^{E_N = E_0} \Delta E_j K_{\gamma  \to e^+} (E_j, E_i) F_{\gnt} (E_j)
\nn\\
&\quad +\sum_{E_j>E_i}^{E_N = E_0} \Delta E_j K_{e^+  \to e^+} (E_j, E_i) F_{\epnt} (E_j)
\Bigg]\,,
\eal
where $\Delta E_j = (E_{j+1} - E_{j-1})/2$.

Photons that maintain energy higher than nuclei's photo-dissociation thresholds can dissociate light elements such as $\D$, $\Be$, etc., which is required for the second part of our bipartite solution (recall Fig.\,\ref{fig:schematic}).
However, they lose their energy via $\gnt+ \gamma \to e^++e^-$ rapidly if the energy is higher than $m_e^2/20 T$.
Thus, we only consider the injection time $\gtrsim 10^5\,\sec$ (at temperature $\lesssim 10\,\keV$), where the nuclear reaction chain has already been turned off, $\Gamma_A^{\rm SBBN} \simeq 0$ in Eq.\,\eqref{eq:BE_X_J}.
When the injection time is earlier, $f_\gnt$ is extremely suppressed due to rapid thermalization, and thus it has no impact other than the entropy injection to the photon sector.

The Boltzmann equations for the evolution of $\D$, $\Li$, and $\Be$ in the presence of ALP are given by
\bal
    &\dfrac{d X_\D}{d t} \!=\! -X_\D \Gamma_{\D \gamma \to pn} \,,\nn\\
     &\dfrac{d X_{\Li}}{d t} \!=\!  -X_{\Li} \!\!\left(\Gamma_{\Li \gamma \to \Het \He} \!\!+\! \Gamma_{\Li \gamma \to n \Lisix} \!\!+\! \Gamma_{\Li \gamma \to n n p \He}\right)\!,\nn\\
    &\dfrac{d X_{\Be}}{d t} \!=\!-X_{\Be} \Gamma_{\Be \gamma \to \Het \He}\,.
\label{eq:BE_X_phi}
\eal
Restricting $m_\phi$ to be smaller than twice of $\He$ photo-dissociation threshold ($=19.81\,\MeV$ for the $\He + \gamma \to p + \T$ channel), we ignore dissociation of $\He$, and abundances of outgoing particles after $\Li$ and $\Be$ dissociation are negligible.
We do not take into account $\T$ and $^3{\rm He}$ since measurements of primordial $^3{\rm He}$ abundance have a large uncertainty (see, e.g., \cite{2018AJ....156..280B, 2022ApJ...932...60C}) and only an upper bound is meaningful.
Note, however, that if the photon energy were greater than the $\He$ photo-dissociation threshold, $\T$ and $^3{\rm He}$ could be overproduced as remnants of $\He$ photo-dissociation leading to another strong bound.
We do not consider that case as we take $m_\phi \lsim 40\,\MeV$.
The reaction rate $\Gamma_{\gnt j \to k l}$ 
for a process $\gnt j \to k l$ can be written as
\bal
\Gamma_{\gnt j \to k l}
= &\dfrac{\xi_\gamma \sigma_{\gnt j \to k l} (E_0)}{\Gamma_{\gnt} (E_0)}
\nn\\
&+ \int_{E_{\rm min}}^{E_0} \!\! d {E_{\gnt}} F_{\gnt}(E_{\gnt})
\sigma_{\gnt j \to k l} (E_{\gnt})\,,
\label{eq:rate}
\eal
where $\sigma_{\gnt j \to k l} (E_{\gnt})$ is the scattering cross section of $ \gnt j \to k l$. 
In our analysis, we take cross sections from the \texttt{JENDL} library data for the photo dissociation of $\D$\,\cite{Iwamoto03082023} and re-derived the fitting formulae for $\Li$ and $\Be$ cross sections, as summarized in Appendix \ref{sec:x_sections}.

The decay of $\phi$ injects entropy into the photon bath and thus modifies the evolution of the background photon temperature.
The evolution equation is given by
\bal
\frac{\dot T}{T}=\frac{\Gamma_\phi \rho_\phi}{4\rho_\gamma} - H \, ,
\eal
where $\rho_\gamma = 2\frac{\pi^2}{30}T^4$ and $\rho_\phi\simeq m_\phi n_\phi$ are energy densities of the background photon and $\phi$.

\begin{figure}[t]
    \centering
    \includegraphics[width=0.5\textwidth]{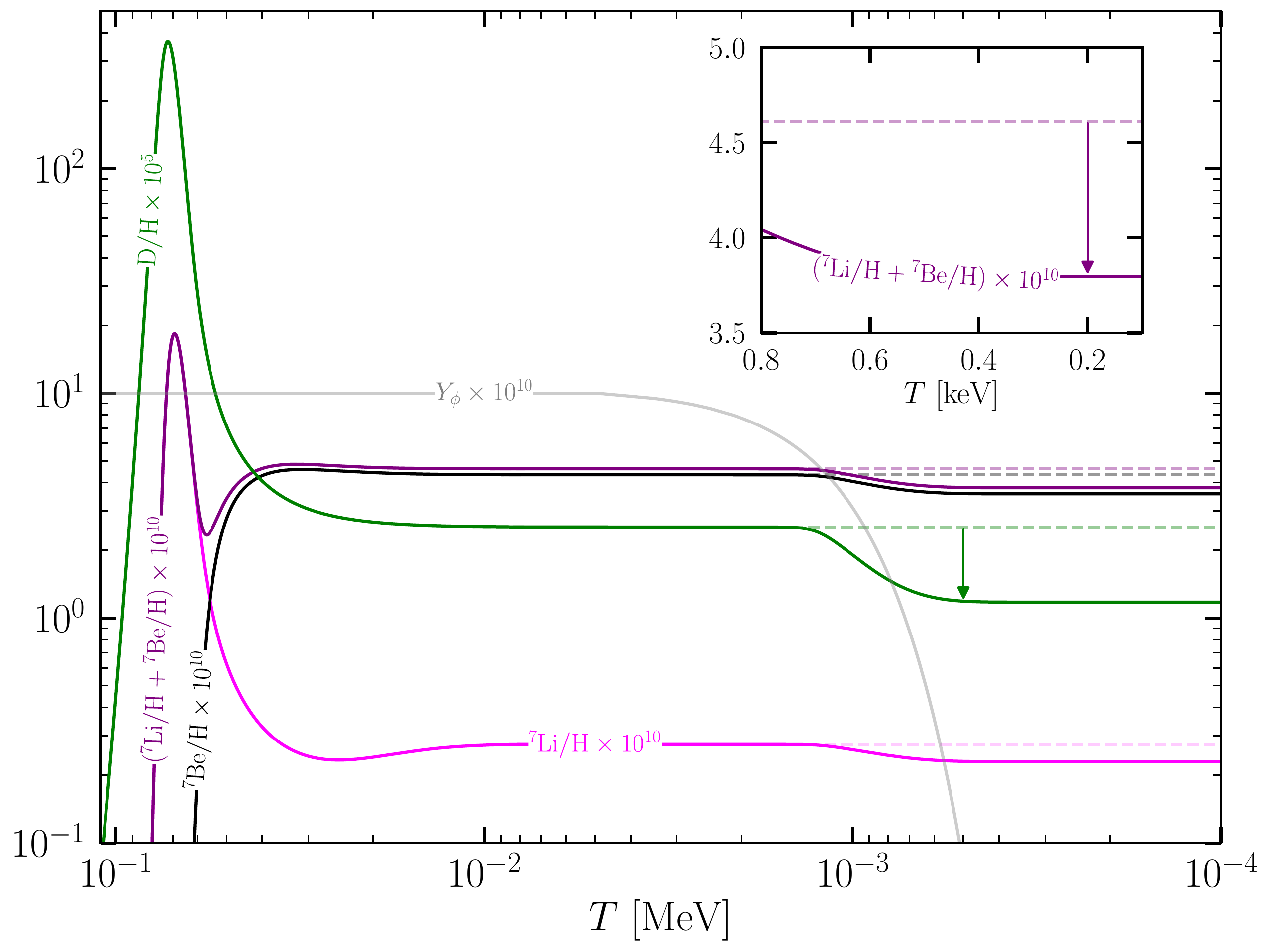}
    \caption{Evolution of 
    $\D/\H$ (green), $\Li/\H$ (magenta),  $\Be/\H$ (black), $(\Li+\!\Be)/\H$ (purple), and $Y_\phi$ (gray) are shown as a function of $T$ for
    $m_\phi = 20\,\rm MeV$, 
    $\tau_\phi = 10^6\,\rm sec$,
    and $Y_\phi^{(0)} = 10^{-9}$. The dashed lines
    depict their evolution
    in SBBN.
    }
    \label{fig:evol_wf}
\end{figure}

In Fig.\,\ref{fig:evol_wf}, we show the evolution of $\D/\H$ (green), $\Li/\H$ (magenta), $\Be/\H$ (black), $(\Li+\!\Be)/\H$ (purple), and
$Y_\phi$ (gray) for the SBBN case (dashed) and the ALP case (solid) with $m_\phi = 20\,\rm MeV$, $Y_\phi^{(0)} = 10^{-9}$, and $\tau_\phi = 10^6\,\rm sec$.
Since the injected energy is larger than the photo-dissociation threshold of $\D$, $\Li$, and $\Be$, nonthermal photons can destroy $\D$, $\Li$, and $\Be$.

\begin{figure}
    \centering
    \includegraphics[width=0.48\textwidth]{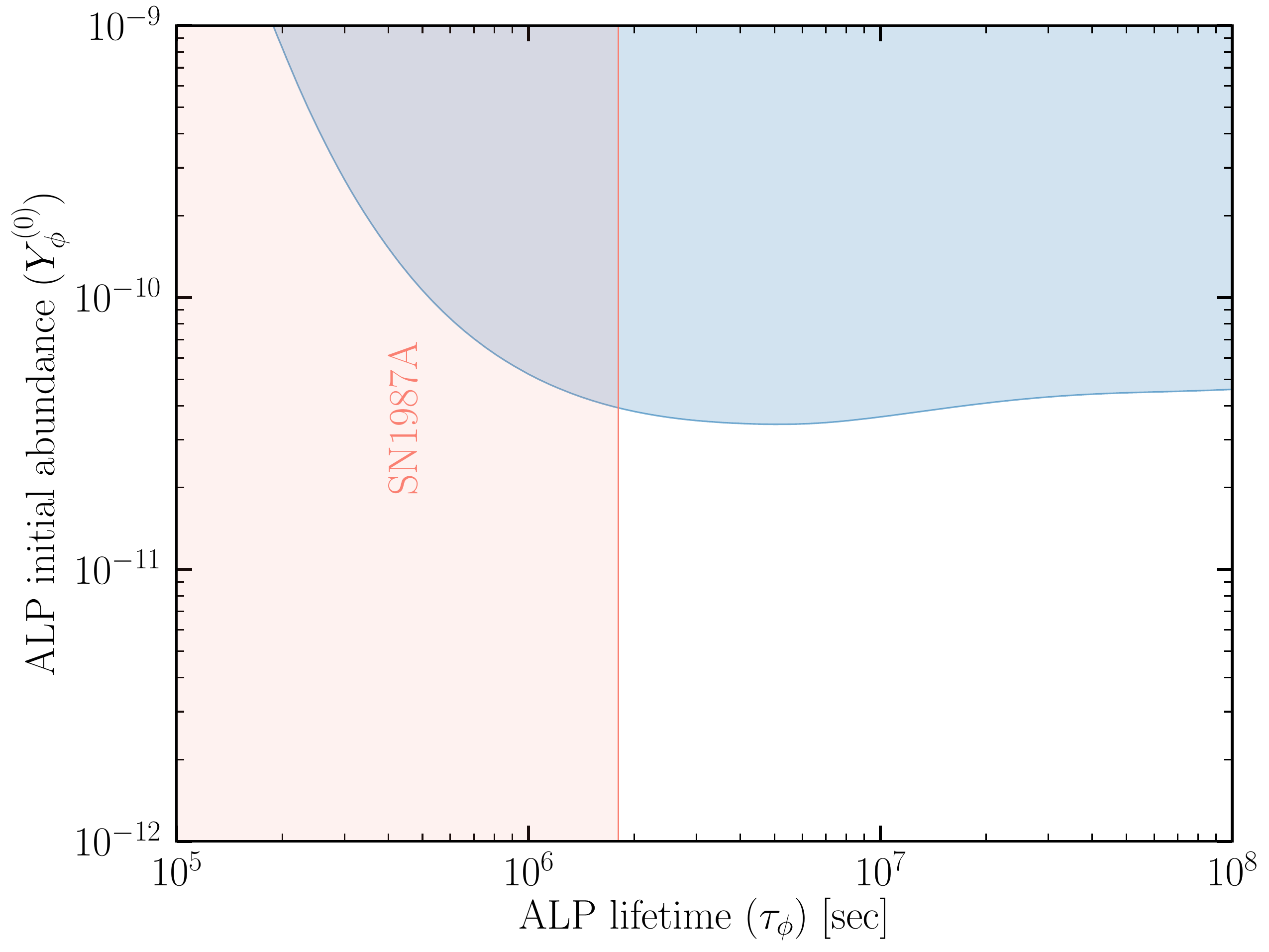}
    \caption{BBN constraint on $\tau_\phi-Y_\phi^{(0)}$ plane for $m_\phi = 20\,\rm MeV$. The light blue region is excluded from the deuterium 
    abundance at 95\% C.L. The orange shaded region depicts the ALP-induced
    gamma-ray burst constraint from SN1987A, and it is taken from
    \cite{AxionLimits}.}
    \label{fig:BBN_exclu_gamma_injection}
\end{figure}

If the first part of our scenario is absent, the deuterium photo-dissociation provides a strong constraint.
In Fig.\,\ref{fig:BBN_exclu_gamma_injection}, we depict our numerical estimation of the constraint (shaded blue region) for $m_\phi = 20\,\rm MeV$ in $\tau_\phi-Y_\phi^{(0)}$ plane.
As one can see, the constraint becomes significantly weaker when 
$\tau_\phi < 10^6\,\rm sec$ due to the rapid thermalization of injected photons.
For \(\tau_\phi \gtrsim 10^6\), photons are essentially collision-free, and most of the emitted photons can efficiently induce photodissociation. Consequently, the bound on the initial abundance shows only a weak dependence on \(\tau_\phi\), becoming nearly flat in this regime. The orange-shaded region indicates the 
ALP-induced gamma-ray burst constraint from SN1987A \cite{Caputo:2021rux, Hoof:2022xbe, Muller:2023vjm}.

\section{SOLUTION TO THE LITHIUM PROBLEM}
\label{sec:results}

Let us now combine the majoron part and the ALP part together, and show how our bipartite solution works.
We first obtain the light nuclei abundances in the majoron part, and they are taken as a boundary condition for the ALP part to obtain their final abundances.

\begin{figure}
    \centering
    \includegraphics[width=0.48\textwidth]{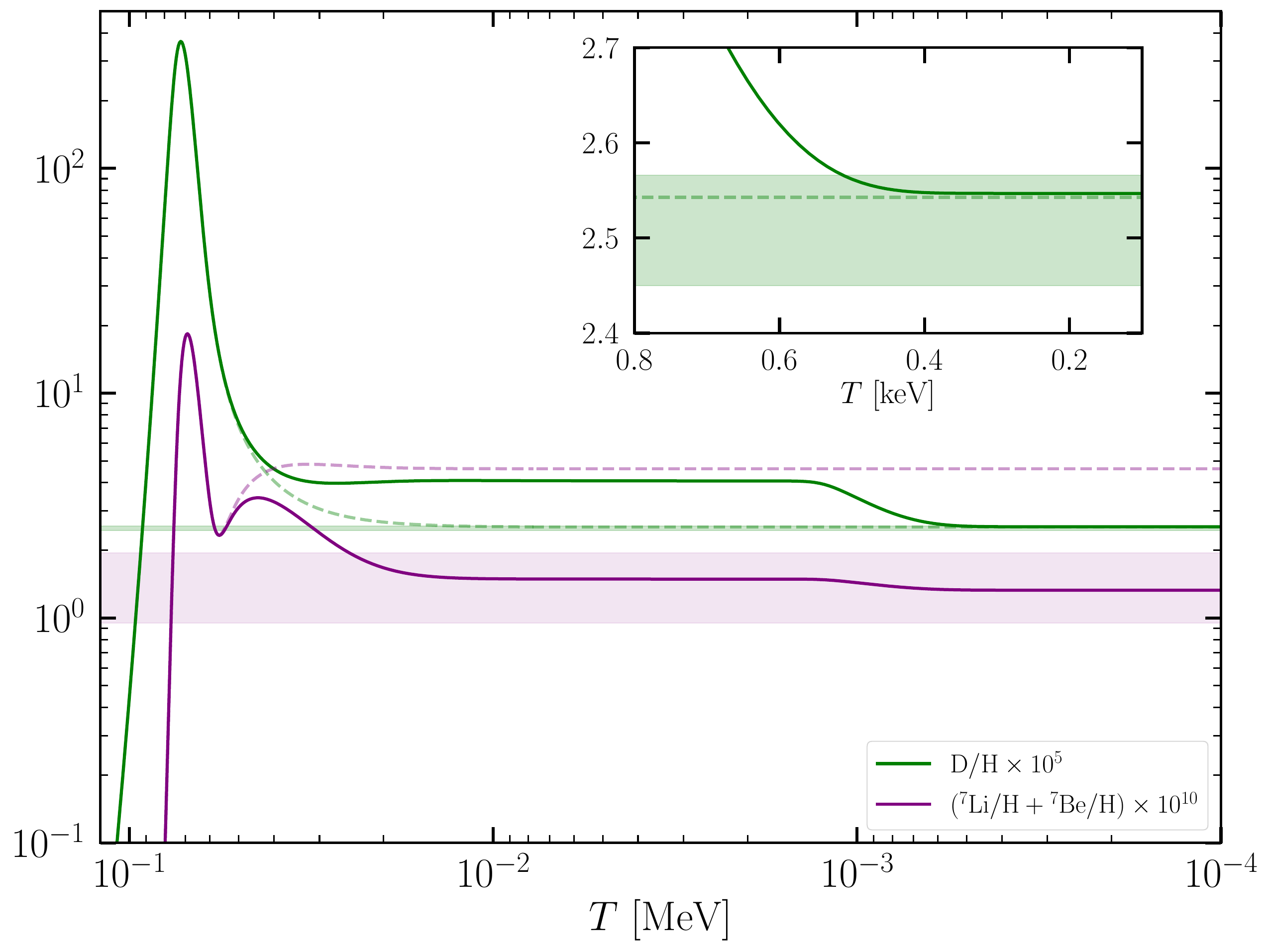}
    \caption{Evolution of $\D/\H$ (green) and $\Li/\H + \!\Be/\H$ (purple) as a function of $T$. The dashed lines depict their evolution for the SBBN scenario. 
    The observational limits of $\D/\H$ and $(\Li/\H + \!\Be/\H)$ at 95\% C.L. are shown by green
    and purple bands, respectively. Here, we choose 
    $(m_J,\,\tau_J,\, Y_J^{(0)}) = (100\,\rm MeV, 10^3\,\rm sec, \,3 \times 10^{-6})$, and 
    $(m_\phi,\,\tau_\phi,\, Y_\phi^{(0)}) = (20\,\rm MeV, 10^6\,\rm sec,\, 6.1 \times 10^{-10})$ 
    to solve the lithium problem while satisfying the constraint on $\D/\H$ abundance.
    }
    \label{fig:evol_wALPandJ}
\end{figure}

In Fig.\,\ref{fig:evol_wALPandJ}, we show the evolution of $\D/\H$, and $(\Li + \Be)/\H$ in temperature $T$, taking benchmark parameters $(m_J,\,\tau_J, Y_J^{(0)}) = (100\,\rm MeV, 10^3\,\rm sec, 3 \times 10^{-6})$, and $(m_\phi,\,\tau_\phi, Y_\phi^{(0)}) = (20\,\rm MeV, 10^6\,\rm sec, 6.1 \times 10^{-10})$.
These parameters are selected to solve the cosmological lithium problem, while satisfying the constraint on $\D/\H$.
As one can see from the figure, $(\Li+\!\Be )/\H$ is reduced and $\D/\H$ is enhanced by the decay of majoron at temperature around $100$ -- $10\,\keV$.
Then, later-time decay of ALP counterbalances the excess $\D$ from the majoron decay, and $\D/\H$ gets back to the $2\sigma$ band of $\D/\H$.
$(\Li/\H + \Be/\H)$ is also further reduced by the ALP, but remains within the $2\sigma$ band.

Scanning the whole parameter space is technically challenging because we have in total six free parameters: $m_J$, $\tau_J$, $Y_J^{(0)}\!$, $m_\phi$, $\tau_\phi$, and $Y_\phi^{(0)}\!$.
Thus, we coarsely scan the parameter space and fit the numerical result with analytic formulae of $\D/\H$ and $\Li/\H$.
The effect from the majoron decay on $\D$ and $\Li$ can be approximated linearly in $Y_J^{(0)}$ since their overproduction is additive. 
We numerically checked that including a quadratic order does not improve the goodness of fitting a lot.
On the other hand, majoron's effect on $\Be$ should be treated as an exponential suppression because destruction always requires itself.
Similarly, the photo-dissociation effects from $Y_\phi^{(0)}$ in both $\D$ and $\Li$ abundances can be treated as an exponential suppression form in $Y_\phi$.
This physical picture motivates the following fitting ansatz for a given $\tau_J$ and $\tau_\phi$:
\bal
\left(\D/\H\right)_{\rm fin.} &= \left(a_1 + b_1 Y_J^{(0)}\right) \exp\left(-c_1 Y_\phi^{(0)}\right),
\label{eq:DbyH_fit}
\\
\left(\Li/\H\right)_{\rm fin.} &= a_2 \left(\exp\left(-b_2 Y_J^{(0)}
\right) +  c_2 Y_J^{(0)}\right) \exp\left(-d_2 Y_\phi^{(0)}\right).\label{eq:LibyH_fit}
\eal
Here, the subscript ``fin" stands for the value estimated at the current Universe and fitting parameters $a_i$, $b_i$, $c_i$, and $d_i$ depend on both mass ($m_J$, $m_\phi$) and lifetime ($\tau_J$, $\tau_\phi$) in general.
Choosing $m_J =100\,\MeV$ and $m_\phi=20\,\MeV$ (which are taken as a benchmark hereafter since dependence on masses is mild) and taking numerical results in $10^{-7} \le Y_J^{(0)} \le 10^{-5}$,
$10^{-10} \le Y_\phi^{(0)} \le 10^{-7}$, 
$10\,{\rm sec} \le \tau_J \le 10^{5}\,\rm sec$,
$10^5\,{\rm sec} \le \tau_\phi \le 10^{8}\,\rm sec$, we obtain fitting parameters summarized in Table\,\ref{tab:tab1} in Appendix\,\ref{sec:fiiting}.
We find that $b_1$, $b_2$, and $c_2$ strongly depend on $\tau_J$, but are not sensitive to $\tau_\phi$.
Similarly, $c_1$ and $d_2$ are sensitive to $\tau_\phi$, but remain insensitive to $\tau_J$. 
When $\tau_\phi \gsim 10^6\,\sec$, $c_1$ and $d_2$ become almost independent of $\tau_\phi$.

\begin{figure*}
    \centering
    \includegraphics[width=0.48\textwidth]{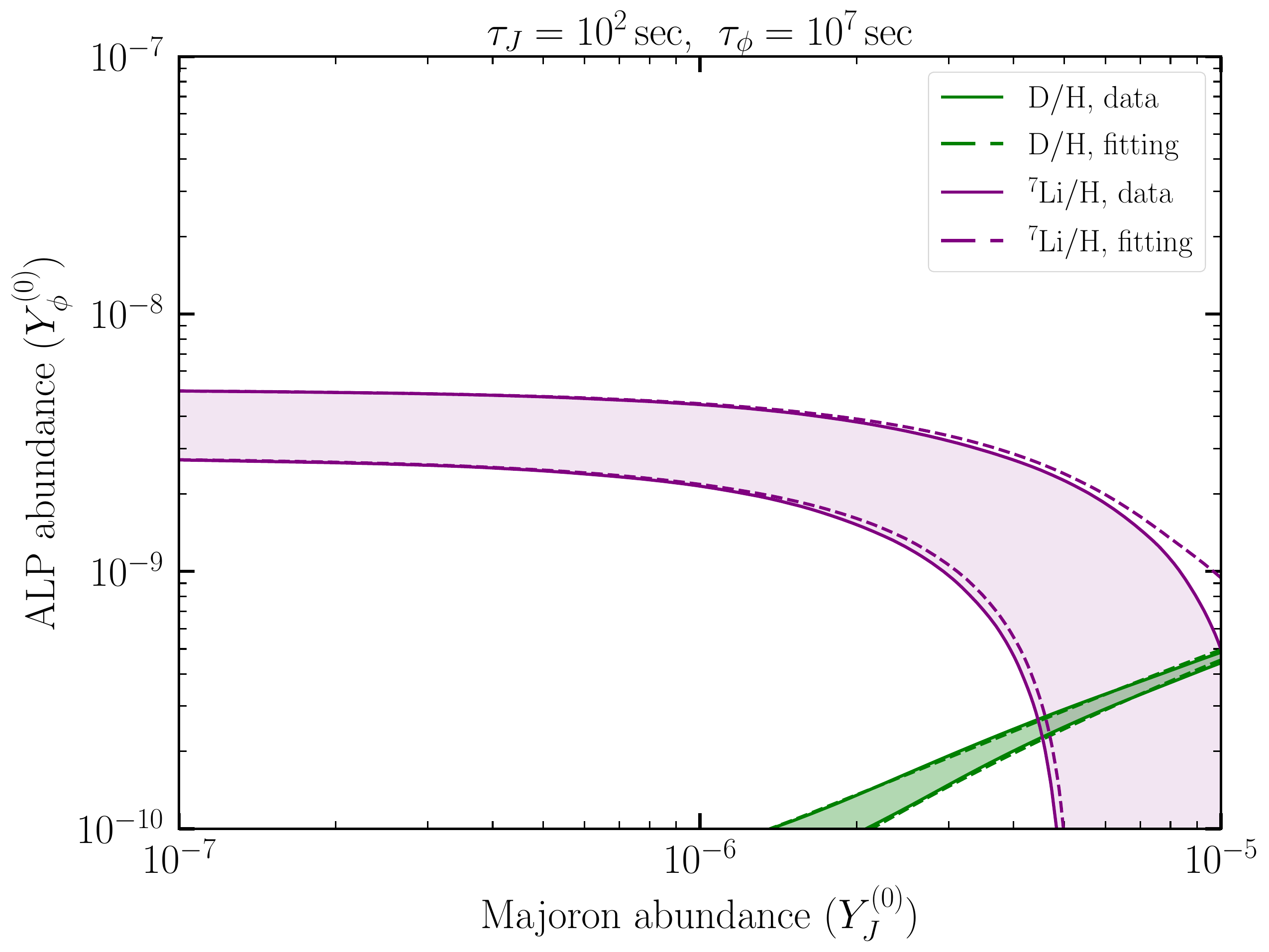}
    \includegraphics[width=0.48\textwidth]{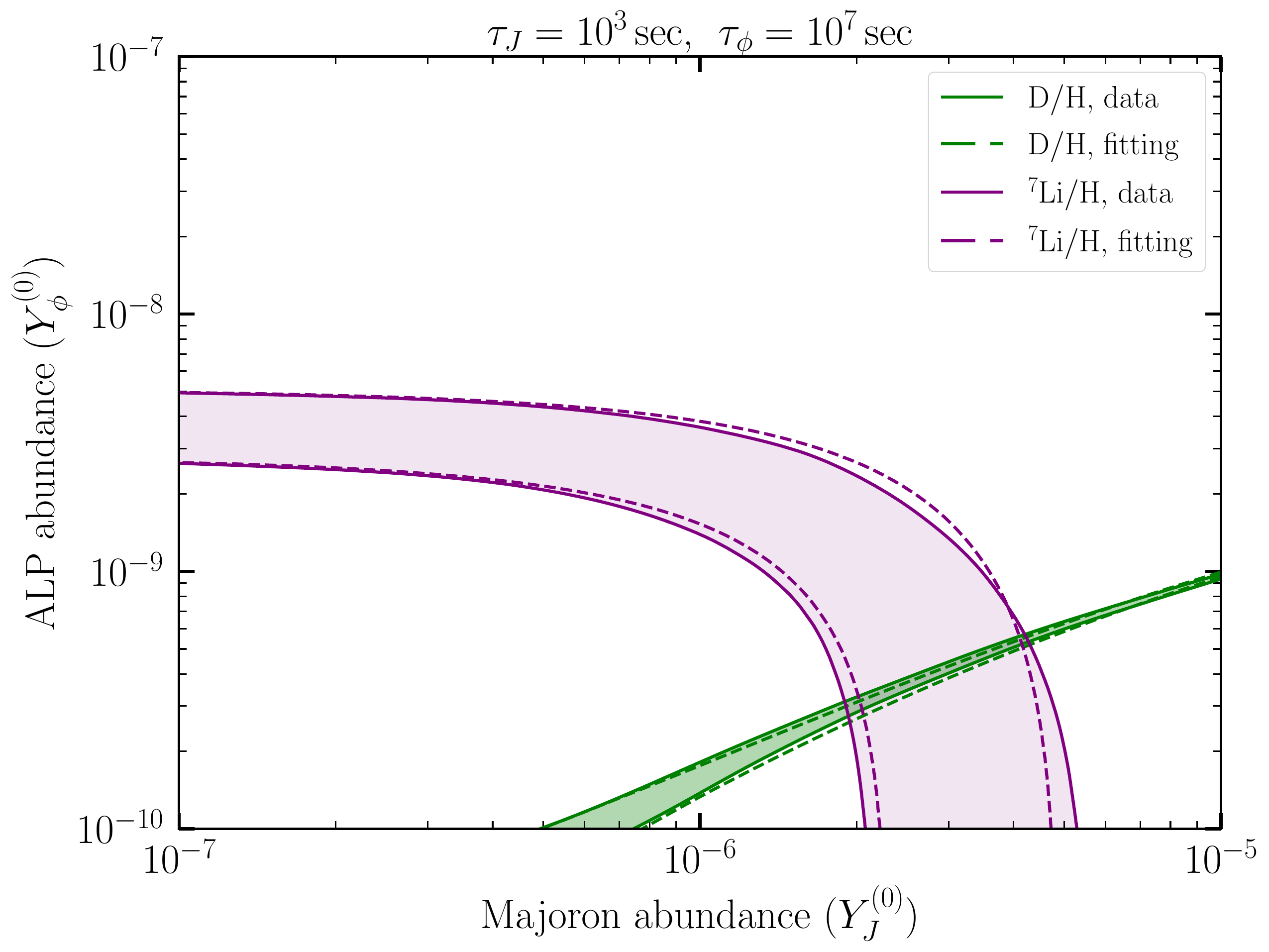}
    \caption{Parameter space in $Y_J^{(0)}-Y_\phi^{(0)}$ plane to explain the observed $\Li$ (purple)
    and $\D$ (green) within 95\% C.L. for $(\tau_J,\,\tau_\phi) = (10^2\,{\rm sec},\, 10^7{\rm sec})$ (left panel) and $(10^3\,{\rm sec},\, 10^7{\rm sec})$ (right panel). 
    In both panels, we consider  $m_J = 100\,\rm MeV$ and $m_\phi = 20\,\rm MeV$.
    Solid contours represent our numerical data, whereas dashed contours correspond to fitting formulae \eqref{eq:DbyH_fit} and \eqref{eq:LibyH_fit} using the parameters in Table.\,\ref{tab:tab1}.
    }
    \label{fig:data_vs_fitting}
\end{figure*}

In Fig.\,\ref{fig:data_vs_fitting} we fix $\tau_J$ and $\tau_\phi$, and find parameter region in $Y_J^{(0)}-Y_\phi^{(0)}$ plane where $\Li$ (purple) and $\D$ (green) abundances can fit within $2\sigma$ range.
Thus, the overlapped region is where the lithium problem is solved.
We take $\tau_J=10^2\,{\rm sec}$, $\tau_\phi = 10^7\,{\rm sec}$ in the left panel and $\tau_J=10^3\,{\rm sec}$, $\tau_\phi = 10^7\,{\rm sec}$ in the right panel. 
In both panels, the contours derived from the numerical data are shown by the solid lines, whereas those derived from our fitting formulae, given in Eq.\,\eqref{eq:DbyH_fit} and \eqref{eq:LibyH_fit} are shown by the dashed lines.
As illustrated in the figure, the parameter space derived from the fitting formulae shows good agreement with that obtained from the numerical data. As indicated by the narrowness of the overlapped region, we need $\sim 10\,\%$ tuning of $Y_\phi^{(0)}$ for an $O(1)$ range of $Y_J^{(0)}$.

\begin{figure}
    \centering
    \includegraphics[width=0.48\textwidth]{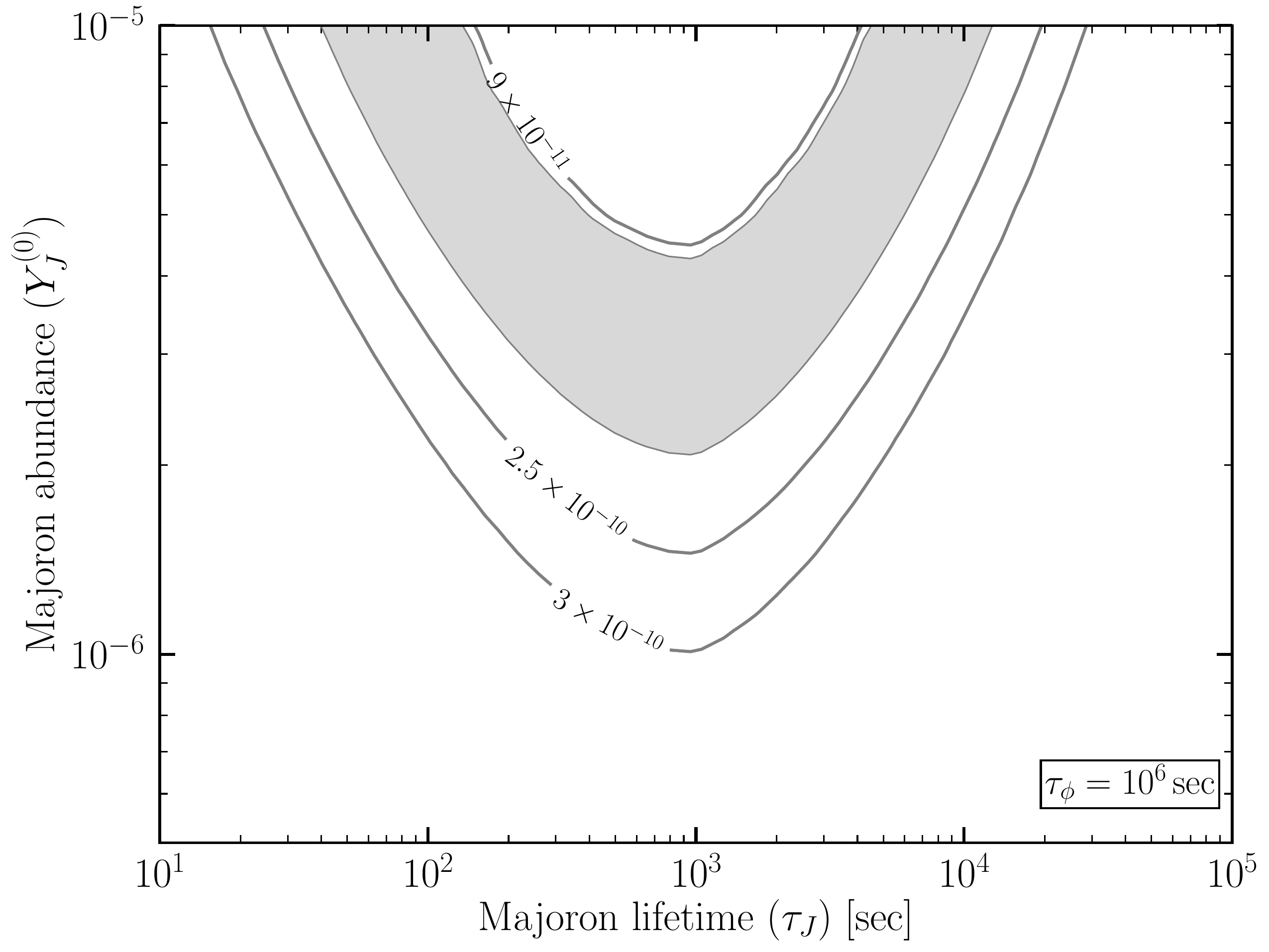}
    \caption{Working parameter space in $\tau_J - Y_J^{(0)}$ plane for $\tau_\phi = 10^6\,\rm sec$ to solve the lithium problem with $m_J=100\,\MeV$ and $m_\phi=20\,\MeV$. $Y_\phi^{(0)}$ is chosen to satisfy the deuterium constraint at each point.
    $(\Li+\!\Be)/\H$ is depicted by gray contours with labels indicating its value, and the shaded gray region solves the lithium problem within $2\,\sigma$.
    }
    \label{fig:taujvsyj}
\end{figure}

\begin{figure}
    \centering
    \includegraphics[width=0.48\textwidth]{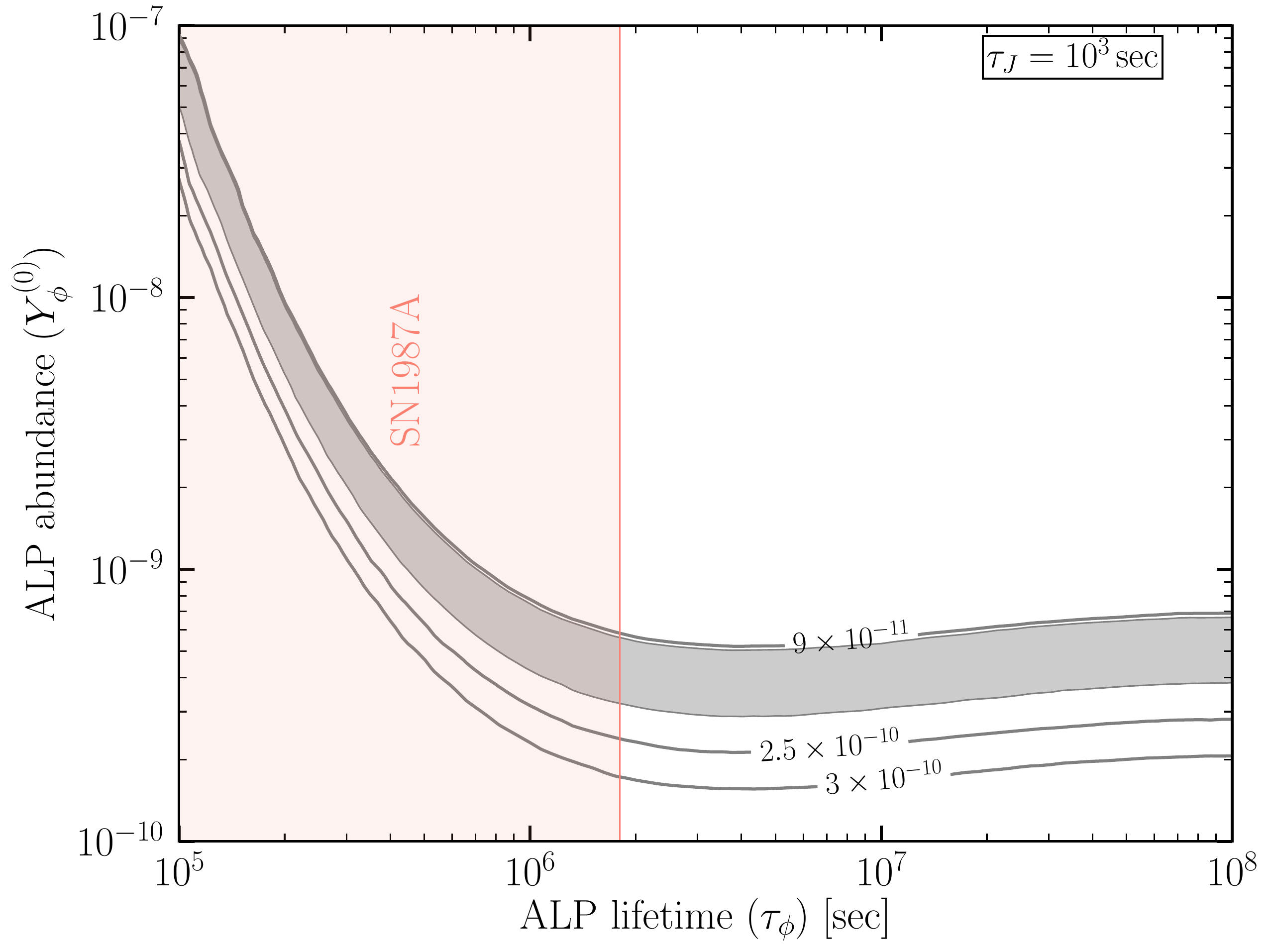}
    \caption{Working parameter space in $\tau_\phi - Y_\phi^{(0)}$ plane for $\tau_J = 10^3\,\rm sec$ to solve the lithium problem with $m_J=100\,\MeV$ and $m_\phi=20\,\MeV$. 
    $Y_J^{(0)}$ is chosen to satisfy the deuterium constraint at each point.
    $(\Li+\!\Be)/\H$ is depicted by gray contours with labels indicating its value, and the shaded gray region solves the lithium problem within $2\,\sigma$. 
    The ALP-induced gamma-ray burst constraint from SN1987A is depicted by the orange shaded region.
    }
    \label{fig:taufvsyf}
\end{figure}

To see the $\tau_J$ dependence of our working parameter space, in Fig.\,\ref{fig:taujvsyj}, we fix $\tau_\phi=10^6\,\sec$, tune $Y_\phi^{(0)}$ to give the observed central value of $\D/\H$, and depict $\Li/\H$ by gray contours.
The shaded region is where the lithium problem is solved up to $2\sigma$.
As expected in the Sec.\,\ref{sec:Majoron_decay}, a wide range of $\tau_J$ is allowed; $10\,\sec \lesssim \tau_J \lesssim 10^4\,\sec$.

Similarly, in Fig.\,\ref{fig:taufvsyf}, we take $Y_J^{(0)}$ tuned for $\D/\H$ with $\tau_J=10^3\,\sec$, and show $\Li/\H$ by gray contours with shaded region representing the solution to the lithium problem up to $2\sigma$.
For $\tau_\phi \lesssim 10^6\,\rm sec$, the required value of $Y_\phi^{(0)}$ to solve the lithium problem is significantly large due to the rapid thermalization of injected photons.
It becomes insensitive to $\tau_\phi$ if $\tau_\phi \gtrsim 10^6\,\rm sec$.

\begin{figure}
    \centering
    \includegraphics[width=0.48\textwidth]{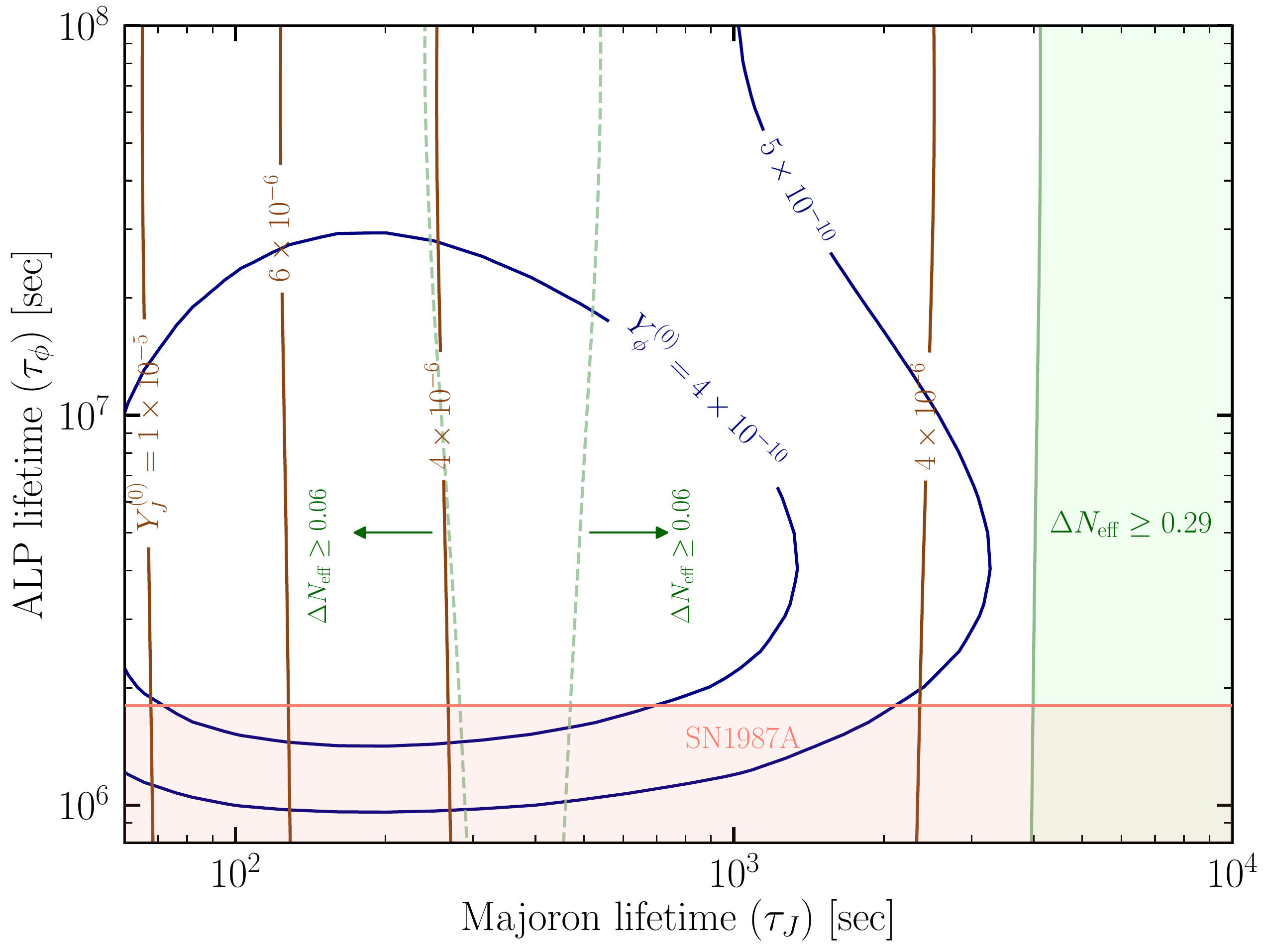}
    \caption{Parameter space in $\tau_J-\tau_\phi$ plane for solving the lithium problem with $m_J = 100\,\rm MeV$ and $m_\phi = 20\,\rm MeV$. 
    $Y_J^{(0)}$ and $Y_\phi^{(0)}$ are chosen to satisfy the observed central values of $\D/\H$ and $(\Li+\!\Be)/\H$ at each point.
    Contours depicted by brown and blue solid lines correspond respectively to $Y_J^{(0)}$ and $Y_\phi^{(0)}$ with labels indicating their values. 
    $\Delta N_{\rm eff}$ constraint from Planck 2018 data is shown by the light green shaded region, whereas the light green dashed lines denote future sensitivity expected when the CMB-S4 experiment takes place.
    The orange shaded region depicts the ALP-induced gamma-ray burst constraint from SN1987A.}
    \label{fig:tauJ_vs_taufi}
\end{figure}

As a final result, for each $\tau_J$ and $\tau_\phi$, we find best fit values of $Y_J^{(0)}$ and $Y_\phi^{(0)}$ for $\D/\H$ and $\Li/\H$, which are depicted by brown and blue contours in Fig.\,\ref{fig:tauJ_vs_taufi}.
For $\tau_J \gtrsim 4 \times 10^3\,\rm sec$, the parameter space that could potentially solve the lithium problem is excluded by the $\Delta N_{\rm eff}$ constraint from Planck 2018 data\,\cite{Planck:2018vyg} and this exclusion becomes stronger if a future experiment such as CMB-S4\,\cite{CMB-S4:2016ple} can take place.

\section{SUMMARY AND CONCLUSIONS}
\label{sec:conclu}
 
In this work, we have investigated a bipartite solution to the cosmological lithium problem. 
Our scenario consists of two distinct stages that work together to reconcile the predicted light-element abundances with observations. 
In the first part, neutrinos are injected from the decay of the majoron with its lifetime in the range of $10\,\text{sec} \lesssim \tau_J \lesssim 10^4\,\text{sec}$. The injected neutrinos enhance the $p \to n$ conversion rate and thus increase the neutron abundance. These excess neutrons successfully decrease the total $\Li+\!\Be$ abundance, but simultaneously lead to an increase in $\D$.

The overproduced deuterium is then addressed in the second part of our scenario. 
By introducing energetic photons from ALP decay with $\tau_\phi \gtrsim 10^5\,\text{sec}$, the excess $\D$ is reduced through photo-dissociation, while the $\Li+\!\Be$ abundance is further decreased. 
We find that this mechanism can reconcile both lithium and deuterium abundances with the observational data.

It is essential for our scenario that $\tau_J$ and $\tau_\phi$ lie in the appropriate ranges, corresponding to the timing of the relevant nuclear and electromagnetic processes. 
For the majoron to be effective, $\tau_J$ should be long enough for substantial $\Be$ to be converted to $\Li$ via $\Be(n, p) \Li$ with the generated extra neutrons, and short enough that the excess $\Li$ can still be burned through $\Li(p,\alpha)\He$. 
For the ALP to be effective, $\tau_\phi$ should be long enough for the Universe to become sufficiently cooled and transparent to the generated high-energy photons. 

This scenario requires an inherent tuning between the two parts in order to reproduce the observed $\D$ abundance, which we estimate quantitatively to be of order $\sim10\,\%$ in the relation between $Y^{(0)}_\phi$ and $Y_J^{(0)}$. 
With this requirement, the bipartite approach provides a physically consistent framework for resolving the lithium problem within the main cosmological constraints relevant to this setup.

The parameter region explored in our analysis is therefore not arbitrary, but is guided by the known response of light-element abundances to different decay channels and decay epochs. At the same time, the small initial yields required for a successful bipartite solution are sensitive to the production history of $J$ and $\phi$, and may point to nonthermal production or a low reheating temperature rather than a minimal thermal origin. Our setup should therefore be viewed as an explicit proof of concept and as a delineation of viable target regions for future model building, rather than as a complete UV-motivated realization. With this caveat, the present scenario still suggests that, once the deuterium constraint is imposed at its current precision, a viable solution to the lithium problem may require a correlated decay history rather than a single late-time modification.

\begin{acknowledgments}
This work was supported by IBS, under the project codes IBS-R018-D1 and IBS-R031-D1. CSS is supported by the National Research Foundation of Korea grant funded by the Korea government RS-2025-25442707 and RS-2026-25498521.
\end{acknowledgments}

\begin{widetext}
\appendix
\section{PHOTO-DISSOCIATION CROSS SECTIONS OF $\Li$ AND $\Be$}
\label{sec:x_sections}
In this appendix, we provide the photodissociation cross sections of $\Be$, $\Li$, and $\D$.
For the processes not discussed here, we adopt the expressions given in Ref.~\cite{Cyburt:2002uv},which we find to be consistent with the experimental data.
Here the energy of the injected photons in unit of MeV is denoted by $\bar{E}_{\gamma}$. The photodissociation
cross-sections are as follows.
\begin{enumerate}
\item $\underline{\Be (\gamma,\,\Het) \He}:$
\bal
\sigma_{\Be \gamma \to \Het \He} = 
\dfrac{0.5}{1.43} \left(\dfrac{\mu}{\rm MeV}\right) \bar{E}_{\rm cm} \bar{E}_{\gamma}^{-2}
\left(\dfrac{\exp\left(-2 \pi \eta\right) }{\bar{E}_{\rm cm}}\right)
{\cal G}_1 (\bar{E}_{\rm cm})\,\,,
\eal
where the threshold energy in unit of MeV $(\bar{Q})$ = 1.587,  $\bar{E}_{\rm cm} = \bar{E}_\gamma - \bar{Q}$, 
$2 \pi \eta = 5.191/\sqrt{\bar{E}_{\rm cm}}$, 
$\mu = 1602.16\,\rm MeV$ and
the functional form of ${\cal G}_1 (\bar{E}_{\rm cm})$ is given by
\bal
{\cal G}_1 (\bar{E}_{\rm cm}) 
&= 0.493\,{\rm mb} \exp\left(-0.568 \bar{E}_{\rm cm}\right)
\left[1 + 0.185 \bar{E}_{\rm cm}^2 + 
0.04 \bar{E}_{\rm cm}^3
+0.015 \bar{E}_{\rm cm}^4
\right]\,\,.
\eal
\item $\underline{\Li (\gamma,\,\T) \He}:$
\bal
\sigma_{\Li \gamma \to \He \T}&=
(0.01813\ \text{mb})\ e^{-0.5607 \bar{E}_\gamma} \times
(1 - 18.52 \bar{E}_\gamma^2 + 6.748 \bar{E}_\gamma^3 + 0.1811 \bar{E}_\gamma^4) .
\eal
where $\bar{Q}=2.46703$ is the threshold energy of the reaction, in units of MeV.
\item $\underline{\Li (\gamma,\,n) \Lisix}:$
\bal
\sigma_{\Li \gamma \to n \Lisix}&=
0.176\, {\rm mb} \,\bar{Q}^{1.51} \bar{E}_{\rm cm}^{0.49} \bar{E}_\gamma^{-2}
+ 1205\, {\rm mb} \,\bar{Q}^{5.5} \bar{E}_{\rm cm}^{5} \bar{E}_\gamma^{-10.5}
+ 0.06\, {\rm mb}\left[1 + \left(\dfrac{\bar{E}_{\rm cm} - 7.46}{0.188}\right)^2\right]^{-1}\,\,,
\eal
where $\bar{Q}= 7.25$, and $\bar{E}_{\rm cm} = \bar{E}_\gamma - \bar{Q}$.
\item $\underline{\Li (\gamma,\,2n\,p) \He}:$
\bal
\sigma_{\Li \gamma \to 2n p \He}&=
122\,{\rm mb} \, \dfrac{ \bar{Q}^4\, \bar{E}_{\rm cm}^3}{\bar{E}_\gamma^7}\,\,,
\eal
where $\bar{Q} = 10.95\,\rm MeV$, and $\bar{E}_{\rm cm} = \bar{E}_\gamma - \bar{Q}$.
\end{enumerate}

\section{FITTING COEFFICIENTS}
\label{sec:fiiting}
The fitting coefficients in Eq.\,\eqref{eq:DbyH_fit}, and \eqref{eq:LibyH_fit} are given in Tab.\,\ref{tab:tab1}.
\begin{table*}[t!]
    \centering
    \begin{tabular}{||c|| c | c | c || c | c | c | c ||}
    \hline
    \multicolumn{8}{||c||}{$\tau_\phi = 10^5\,\rm sec$}\\
    \hline
     $\tau_J\,[\rm sec]$ &   $a_1$ &  $b_1$ & $c_1$ & $a_2$ &$b_2$ &$c_2$ & $d_2$\\
     \hline
     $10^1$ & $2.52 \times 10^{-5}$  & $0.38$  & $6.64 \times 10^6$      & $4.64 \times 10^{-10}$ &  $4.78 \times 10^4$  & $5.43 \times 10^3$  &  $1.27 \times 10^6$ \\
     $10^2$ & $2.54 \times 10^{-5}$  & $1.67$  & $6.64\times 10^6$       & $4.63 \times 10^{-10}$ &  $2.04 \times 10^5$  & $1.15 \times 10^4$  &  $1.25 \times 10^6$\\
     $10^3$ & $2.57 \times 10^{-5}$  & $4.60$  & $6.64\times 10^6$       & $4.59 \times 10^{-10}$ &  $4.11 \times 10^5$  & $1.44 \times 10^4$  &  $1.21 \times 10^6$\\
     $10^4$ & $2.53 \times 10^{-5}$  & $2.67$  & $6.64\times 10^6$       & $4.63 \times 10^{-10}$ &  $1.24 \times 10^5$  & $1.22 \times 10^4$  &  $1.16 \times 10^6$\\
     $10^5$ & $2.52 \times 10^{-5}$  & $0.52$  & $6.64\times 10^6$       & $4.64 \times 10^{-10}$ &  $1.90 \times 10^4$  & $2.38 \times 10^3$  &  $1.22 \times 10^6$\\
    \hline

    \multicolumn{8}{||c||}{$\tau_\phi = 3.98 \times 10^5\,\rm sec$}\\
    \hline
     $10^1$ & $2.52 \times 10^{-5}$ & $0.38$ & $2.77\times 10^8$      & $4.64 \times 10^{-10}$  &  $4.78 \times 10^4$  & $5.44 \times 10^3$  &  $6.61 \times 10^7$ \\
     $10^2$ & $2.54 \times 10^{-5}$ & $1.67$ & $2.77\times 10^8$      & $4.63 \times 10^{-10}$  &  $2.04 \times 10^5$  & $1.15 \times 10^4$  &  $6.59 \times 10^7$\\
     $10^3$ & $2.57 \times 10^{-5}$ & $4.60$ & $2.77\times 10^8$      & $4.59 \times 10^{-10}$  &  $4.10 \times 10^5$  & $1.45 \times 10^4$  &  $6.53 \times 10^7$\\
     $10^4$ & $2.53 \times 10^{-5}$ & $2.67$ & $2.77\times 10^8$      & $4.62 \times 10^{-10}$  &  $1.22 \times 10^5$  & $1.21 \times 10^4$  &  $6.45 \times 10^7$\\
     $10^5$ & $2.52 \times 10^{-5}$ & $0.52$ & $2.77\times 10^8$      & $4.64 \times 10^{-10}$  &  $1.86 \times 10^4$  & $2.21 \times 10^3$  &  $6.55 \times 10^7$\\
    \hline

    \multicolumn{8}{||c||}{$\tau_\phi = 1.58 \times 10^6\,\rm sec$}\\
    \hline
     $10^1$ & $2.52 \times 10^{-5}$ & $0.38$ & $9.66\times 10^8$    & $4.64 \times 10^{-10}$  &  $4.78 \times 10^4$  & $5.42 \times 10^3$ &  $2.51 \times 10^8$ \\
     $10^2$ & $2.54 \times 10^{-5}$ & $1.67$ & $9.66\times 10^8$    & $4.63 \times 10^{-10}$  &  $2.04 \times 10^5$  & $1.14 \times 10^4$ &  $2.50 \times 10^8$\\
     $10^3$ & $2.57 \times 10^{-5}$ & $4.60$ & $9.66\times 10^8$    & $4.60 \times 10^{-10}$  &  $4.12 \times 10^5$  & $1.43 \times 10^4$ &  $2.50 \times 10^8$\\
     $10^4$ & $2.53 \times 10^{-5}$ & $2.67$ & $9.66\times 10^8$    & $4.64 \times 10^{-10}$  &  $1.25 \times 10^5$  & $1.22 \times 10^4$ &  $2.49 \times 10^8$\\
     $10^5$ & $2.52 \times 10^{-5}$ & $0.52$ & $9.66\times 10^8$    & $4.64 \times 10^{-10}$  &  $1.92 \times 10^4$  & $2.47 \times 10^3$ &  $2.50 \times 10^8$\\
    \hline

    \multicolumn{8}{||c||}{$\tau_\phi = 6.31 \times 10^6\,\rm sec$}\\
    \hline
     $10^1$ & $2.52 \times 10^{-5}$ & $0.38$ & $1.13\times 10^9$    & $4.64 \times 10^{-10}$ &  $4.79 \times 10^4$  & $5.37 \times 10^3$ & $3.20 \times 10^8$\\
     $10^2$ & $2.54 \times 10^{-5}$ & $1.67$ & $1.13\times 10^9$    & $4.63\times 10^{-10}$  &  $2.05 \times 10^5$  & $1.12 \times 10^4$ & $3.21 \times 10^8$\\
     $10^3$ & $2.57 \times 10^{-5}$ & $4.60$ & $1.13\times 10^9$    & $4.61 \times 10^{-10}$ &  $4.17 \times 10^5$  & $1.39 \times 10^4$ & $3.24 \times 10^8$\\
     $10^4$ & $2.53 \times 10^{-5}$ & $2.67$ & $1.13\times 10^9$    & $4.66 \times 10^{-10}$ &  $1.29 \times 10^5$  & $1.24 \times 10^4$ & $3.27 \times 10^8$\\
     $10^5$ & $2.52 \times 10^{-5}$ & $0.52$ & $1.13\times 10^9$    & $4.65 \times 10^{-10}$ &  $2.09 \times 10^4$  & $3.08 \times 10^3$ & $3.23 \times 10^8$\\
     \hline

    \multicolumn{8}{||c||}{$\tau_\phi = 2.51 \times 10^7\,\rm sec$}\\
    \hline
     $10^1$ & $2.52 \times 10^{-5}$ & $0.38$ & $9.25\times 10^8$   & $4.64 \times 10^{-10}$  &  $4.81 \times 10^4$ & $5.33 \times 10^3$  & $2.82 \times 10^8$\\
     $10^2$ & $2.54 \times 10^{-5}$ & $1.67$ & $9.25\times 10^8$   & $4.63 \times 10^{-10}$  &  $2.06 \times 10^5$ & $1.10 \times 10^4$  & $2.84 \times 10^8$\\
     $10^3$ & $2.57 \times 10^{-5}$ & $4.60$ & $9.25\times 10^8$   & $4.62 \times 10^{-10}$  &  $4.21 \times 10^5$ & $1.35 \times 10^5$  & $2.89 \times 10^8$\\
     $10^4$ & $2.53 \times 10^{-5}$ & $2.67$ & $9.25\times 10^8$   & $4.68 \times 10^{-10}$  &  $1.34 \times 10^5$ & $1.26 \times 10^4$  & $2.95 \times 10^8$\\
     $10^5$ & $2.52 \times 10^{-5}$ & $0.52$ & $9.25\times 10^8$   & $4.66 \times 10^{-10}$  &  $2.25 \times 10^4$ & $3.58 \times 10^3$  & $2.87 \times 10^8$\\
     \hline

    \multicolumn{8}{||c||}{$\tau_\phi =  10^8\,\rm sec$}\\
    \hline
     $10^1$ & $2.52 \times 10^{-5}$ & $0.38$ & $8.47\times 10^8$    & $4.64 \times 10^{-10}$ &  $4.81 \times 10^4$ & $5.31 \times 10^3$ & $2.64 \times 10^8$\\
     $10^2$ & $2.54 \times 10^{-5}$ & $1.67$ & $8.47\times 10^8$    & $4.63 \times 10^{-10}$ &  $2.06 \times 10^5$ & $1.09 \times 10^4$ & $2.66 \times 10^8$\\
     $10^3$ & $2.57 \times 10^{-5}$ & $4.60$ & $8.47\times 10^8$    & $4.62 \times 10^{-10}$ &  $4.23 \times 10^5$ & $1.33 \times 10^4$ & $2.72 \times 10^8$\\
     $10^4$ & $2.53 \times 10^{-5}$ & $2.67$ & $8.47\times 10^8$    & $4.68 \times 10^{-10}$ &  $1.36 \times 10^5$ & $1.26 \times 10^4$ & $2.79 \times 10^8$\\
     $10^5$ & $2.52 \times 10^{-5}$ & $0.52$ & $8.47\times 10^8$    & $4.66 \times 10^{-10}$ &  $2.31 \times 10^4$ & $3.75 \times 10^3$ & $2.70 \times 10^8$\\
     \hline
    \end{tabular}
    \caption{Fitting coefficients for $m_J = 100\,\rm MeV$, and $m_\phi = 20\,\rm MeV$.}
    \label{tab:tab1}
\end{table*}
\end{widetext}

\bibliography{refs_v4}

\end{document}